\documentclass[a4paper,12pt]{report}
\usepackage[latin2]{inputenc}
\usepackage{amsfonts}
\usepackage[T1]{fontenc}
\usepackage{graphicx}
\usepackage{amssymb}
\usepackage{amsthm}
\usepackage{amsmath}
\usepackage[small]{caption}
\usepackage{comment}
\usepackage{calc}
\usepackage{hyperref}
\usepackage{color}
\usepackage{xfrac}
\usepackage{tikz}
\usetikzlibrary{decorations.pathmorphing}

\renewcommand{\[}{\begin{equation}}
\renewcommand{\]}{\end{equation}}

\newcommand{\Tr}[1]{\mathrm{Tr}(#1)}

\newcommand{\ket}[1]{|#1\rangle}
\newcommand{\bra}[1]{\langle#1|}
\newcommand{\braket}[2]{\langle#1|#2\rangle}

\definecolor{darkred}{rgb}{.8,0,0}
\definecolor{green}{rgb}{.2,.6,.2}

\newcommand{\smiley}{\tikz[baseline=-0.75ex,black]{
    \draw circle (2mm);
\node[fill,circle,inner sep=0.5pt] (left eye) at (135:0.8mm) {};
\node[fill,circle,inner sep=0.5pt] (right eye) at (45:0.8mm) {};
\draw (-145:0.9mm) arc (-120:-60:1.5mm);
    }
}

\newcommand{\frownie}{\tikz[baseline=-0.75ex,black]{
    \draw circle (2mm);
\node[fill,circle,inner sep=0.5pt] (left eye) at (135:0.8mm) {};
\node[fill,circle,inner sep=0.5pt] (right eye) at (45:0.8mm) {};
\draw (-145:0.9mm) arc (120:60:1.5mm);
    }
}

\newcommand{\neutranie}{\tikz[baseline=-0.75ex,black]{
    \draw circle (2mm);
\node[fill,circle,inner sep=0.5pt] (left eye) at (135:0.8mm) {};
\node[fill,circle,inner sep=0.5pt] (right eye) at (45:0.8mm) {};
\draw (-135:0.9mm) -- (-45:0.9mm);
    }
}

\title{Delayed Choice experiments and causality in Quantum mechanics}
\author{Dominik \v{S}afr\'{a}nek}

\pdfoutput=1

\begin{document}
\hypersetup{pageanchor=false}
\pagestyle{empty}

\begin{center}
{\Large CZECH TECHNICAL UNIVERSITY IN PRAGUE} \\[3mm]
{\Large Faculty of Nuclear Sciences and Physical Engineering}

\vspace{\stretch{0.5}}

{\huge\bf DIPLOMA THESIS}

\vspace{\stretch{1}}

\end{center}
{\Large \hspace*{1cm} 2013 \hfill Dominik \v{S}afr\'{a}nek \hspace*{1cm}}
\newpage
\begin{center}
{\Large CZECH TECHNICAL UNIVERSITY IN PRAGUE} \\[3mm]
{\Large Faculty of Nuclear Sciences and Physical Engineering}

\vspace{\stretch{0.2}}

\begin{figure}[h]
\begin{center}
\includegraphics[scale=0.4]{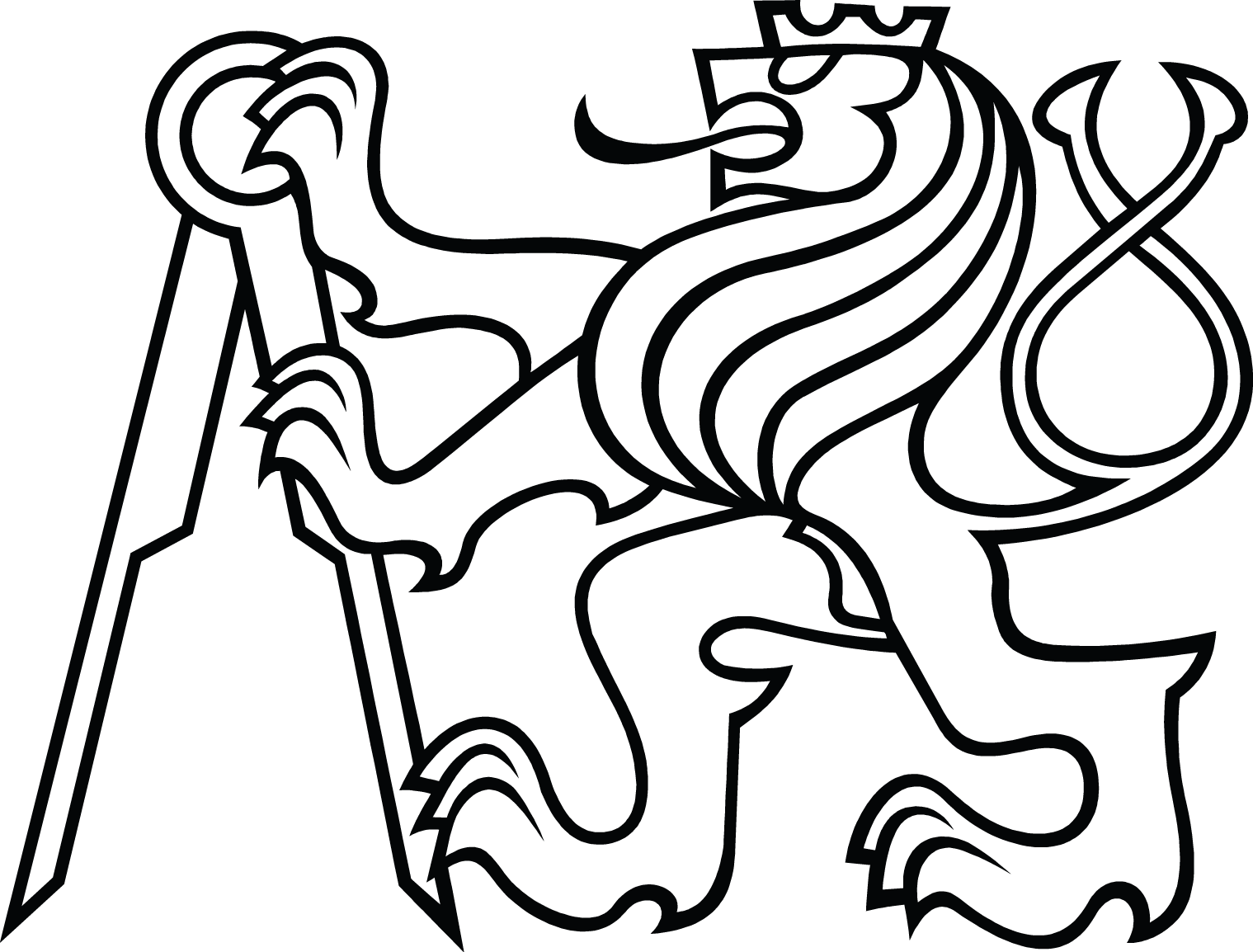}
\end{center}
\end{figure}

\vspace{\stretch{0.2}}

{\huge\bf DIPLOMA THESIS}

\vspace{\stretch{0.1}}

{\huge Delayed Choice experiments}
\vspace{1.3mm}
{\huge {and causality in Quantum mechanics}}
\vspace{\stretch{0.8}}

\end{center}
\begin{tabular}{ll}
\large Author: &\large Dominik \v{S}afr\'{a}nek\\
\large Supervisor: &\large Ing. Petr Jizba, PhD.\\
\large Consultants: &\large Dr. Jacob Dunningham\\
\large Year: &\large 2013\\
\end{tabular}
\newpage
\pagestyle{empty}
\section*{Acknowledgement}

I would like to thank to Va\v sek Poto\v cek for pointing out why Free Will experiment cannot work, to Dr. Jacob Dunningham for extraordinary talks and ideas without which I wouldn't find courage to write what I've been doing, and most of all to Ing. Petr Jizba Ph.D. who had the patience to explain me over and over again, guided me and helped me to realize my dream.

\newpage

\section*{Prohlá\v{s}ení}
Prohla\v{s}uji, \v{z}e jsem svůj výzkumný úkol práci vypracoval samostatně a pou\v{z}il jsem pouze literaturu uvedenou
v p\v{r}ilo\v{z}eném seznamu.

Nemám záva\v{z}ný důvod proti u\v{z}ití \v{s}kolního díla ve smyslu §60 Zákona \v{c} 212/2000 Sb., o právu autorském, o právech
souvisejících s právem autorským a o zm\v{e}n\v{e} n\v{e}kterých zákon\r{u} (autorský zákon).

\section*{Declaration}
I declare, I wrote my Research Project independently and exclusively with the use of cited bibliography.

I agree with the usage of this thesis in the purport of the §60 Act 121/2000 (Copyright
Act).
\\
\\

V Praze dne ...........................\ \ \ \ \ \ \ \ \ \ \ \ \ \ \ \ \ \ \  ......................................
\newpage
\noindent
\begin{tabular}{ll}
\emph{Název práce:}& \textbf{``Delayed choice'' experimenty}\\
 & \textbf{a kauzalita v Kvantové mechanice}\\
\emph{Autor:}& Dominik \v Safránek\\
\emph{Obor:}& Matematické in\v{z}enýrství\\
\emph{Druh práce:}& Výzkumný úkol\\
\emph{Vedoucí práce:}&  Ing. Petr Jizba, PhD. \\
\emph{Konzultanti:}&  Dr. Jacob Dunningham \\
& Katedra fyziky, Fakulta jaderná a fyzikálně in\v{z}enýrská \\
\end{tabular}

\begin{center}\textbf{Abstrakt}\end{center}
\noindent

P\v resto, \v ze se m\r u\v ze zdát, \v ze kvantové experimenty se zpo\v zd\v eným výb\v erem naru\v sují klasickou kauzalitu, (tj. \v ze zdánliv\v e lze sestrojit experiment, který m\r u\v ze ovliv\v novat minulost), pomocí kvantové teorie ``mnoha sv\v et\r u'' dokazujeme, \v ze tomu tak nem\r u\v ze být. Dále matematicky formulujeme koncept ``which-path'' informace a ukazujeme, \v ze její potenciální získatelnost (byt' \v cáste\v cná) zp\r usobuje vyhasnutí interference. Pro lep\v sí názornost konstruujeme my\v slenkový systém, který vykazuje jak interferenci, tak kvantovou korelaci, a demonstrujeme, \v ze interference a korelace jsou nevyhnuteln\v e komplementární koncepty. S pou\v zitím konceptu ``Kvantové gumy'' demonstrujeme, \v ze neexistuje objektivní reality ve smyslu Einsteina, Podolského a Rosena. Dále diskutujeme rozdíl mezi ``vn\v ej\v sím'' (unitárním) a ``vnit\v rním'' (neunitárním) pozorovatelem. Práce je vybavena 3 apendixy, které rozvádí technické detaily této diplomové práce.
\\
\\
\\
\noindent
\begin{tabular}{ll}
\emph{Klí\v cová slova:}& kauzalita, korelace \\
&  which-path informace \\
& experiment se zpo\v zd\v eným výb\v erem\\
& realita, Many World\\
& problém kvantového m\v e\v rení\\
& Kvantová guma
\end{tabular}

\newpage
\noindent
\\
\\
\begin{tabular}{ll}
\emph{Title:}& \textbf{Delayed Choice Experiments }\\
&\textbf{and Particle Entanglement}\\
\emph{Author:}& Dominik \v Safránek\\
\end{tabular}
\\
\begin{center}\textbf{Abstract}\end{center}
\noindent

Although it may seem The Delayed Choice experiments contradict causality and one could construct an experiment which could possibly affect the past, using Many World interpretation we prove it is not possible. We also find a mathematical background to Which-path information and show why its obtainability prevents system from interfering. We find a system which exhibit both interference and correlation and show why one-particle interference and correlations are complementary. Better visible interference pattern leads to worse correlations and vice versa. Then, using knowledge gained from Quantum Eraser and Delayed Choice experiments we prove there is not an objective reality in a sense of Einstein, Podolsky and Rosen. Furthermore, we discuss the difference between ``outer'' (non-interacting) and ``inner'' (interacting) observer. We find the mathematical relationship between the ``universal'' wave function used by ``outer'' observer and processes the ``inner'' observer sees, which is our small contribution to the measurement problem.
\\
\\
\\
\noindent
\begin{tabular}{ll}
\emph{Keywords:}&  causality, correlations \\
& Which-path information \\
& Delayed Choice, Quantum Eraser\\
& reality, Many World interpretation\\
& measurement problem \\
\end{tabular}

\noindent

\tableofcontents
\noindent

\newpage
\hypersetup{pageanchor=true}
\pagestyle{plain}
\pagenumbering{arabic}
\setcounter{page}{1}
\chapter*{Preface}\addcontentsline{toc}{chapter}{Preface}

This thesis is a result of my 3-years long research on the foundations of Quantum Mechanics. I was always interested in fundamental issues and wonderful ideas that enter the realm of the quantum world. In the presented work I have strived to explain some of the pressing issues of contemporary Quantum Mechanics in a simple and possibly non-confusing manner. After exploring the Delayed Choice and Quantum Eraser experiments I started to think what is it all about, what happens when we generalize the knowledge we gained. When you want to understand properly, take it to the extreme, do not fear the results you can get, do not fear to talk about it. When you find the slightest inconsistency, do not move away from it and try to solve it. That is why I introduced the concept of causality, tests of free will, the idea of brain-washing and relative realities. Yet, throughout this thesis I have always tried to support my statements and suggestions by rigorous mathematics or at least by ``thought'' experiments."

\chapter*{Introduction}\addcontentsline{toc}{chapter}{Introduction}

In this thesis we will elaborate on inner workings of ``delayed choice quantum eraser''. This will be first worked out in detail on selected examples and then generalized further. After this we will elaborate philosophical consequences which the aforementioned experiments provide. We will try to justify the general idea of The Many World interpretation, still with some reservations of ours. Most of the physicist believe this interpretation cannot be proven directly, probably not even indirectly, yet we will try show you that this approach elegantly solves quantum eraser experiments while avoiding vague and unnatural statement ``possibly obtainable which-path information prevents system from interfering''. This approach also shows that interference and two-particle interference (correlations) complementarity is not just an experimental fact but also a simple result of The Many World and Wigner's friend approach to Quantum Mechanics.

Then we move to even more unbelievable issue, which we will show is again a simple consequence of existence of quantum eraser experiments and assumption the people are made from the very same atoms as the world around us, so they also behave in the very same way, they follow the same laws of physics.

\bigskip
\textbf{Observers follow the same laws of physics. Observers are just another quantum system.}
\bigskip

We show that there is not an objective reality, but each person can see the reality differently. Reality is relative. Not only about how we feel about things but also if these things exists or not -- they may exist for one observer and not to exist for another. Or they may exists for one, but after you erase the information about its existence from his brain, it may even stop to exist again.

We illustrate that ``worlds'' that enter Many World interpretation are not immune from seeing each other --- the fact that is usually believed to be true. Yet, we show that in order to pass between such distinct worlds we need to follow very special rules which are necessary in order to prevent us from experiencing paradoxes.

At the end of the last chapter we also elaborate on the notorious measurement problem. In particular, we suggest a way how and when the measurement is done and we also find the relationship between the universal function (describing both observer and the observed system) and the reality observer sees. To this end we introduce the ``passive Zeno effect'', that is the effect, where the observer does not notice any change, he does not detect anything, but still he destroys a possible interference.

\chapter{Delayed Choice and Quantum Eraser experiments}

Delayed choice experiments were though experiments at first, but lately with many experimental realizations \cite{JaquesExperimentalRealizationDelayedChoice, PeruzzoDelayedChoiceExperimental, KaiserDelayedChoiceExperimental, ZeilingerDelayedChoiceExperimentalSwapping}. It shows that non-locality is an inherent part of quantum particles and asking where exactly the quantum particle remain before the actual measurement is irrelevant. We can say where particle is just when position measurement has been done, but not before and even not after (due to wave-packet spreading).

Quantum Eraser experiments shows that when a measurement device itself is a quantum system, it is possible to cancel possible interference by measurement, but we can regain it when we erase the information from the measuring device. We will see that some of the eraser experiments does not fit this very well, so maybe it is better to refer to these experiments in a more general way as the processes of destroying and regaining interference patterns.

In the last sections we will deal with delayed choice quantum erasers which combine the previous two and the Free Will experiment, which is the type of delayed choice quantum eraser which could possibly determine whether the person has or has not the free will.

\section{Wheeler's delayed choice}\label{sectiondelayedchoice}

\begin{figure}[h]
\begin{center}
\includegraphics[scale=1]{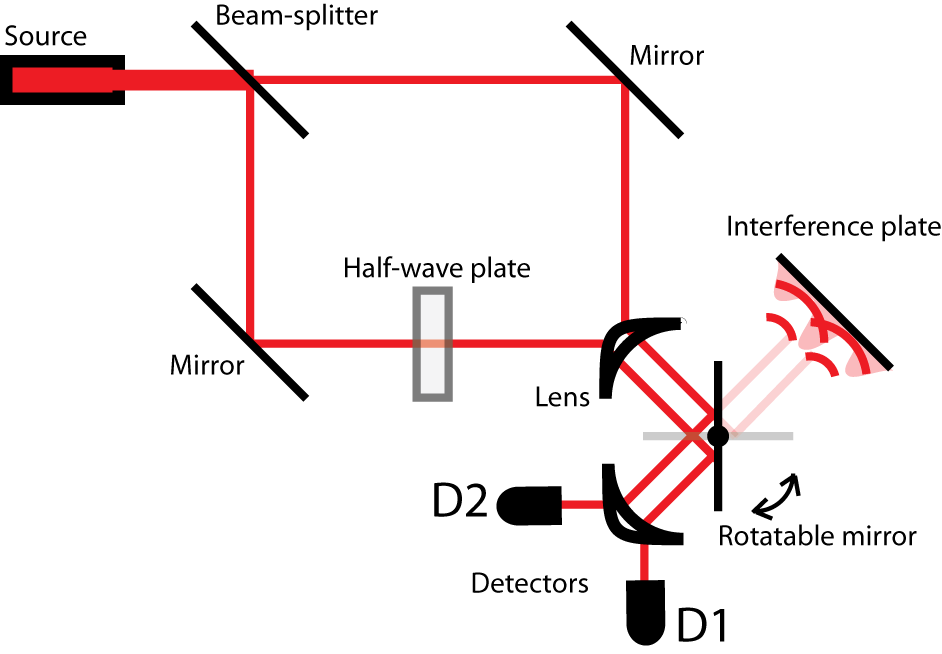}
\caption{Wheeler's delayed choice experiment.}\label{Wheelers}
\end{center}
\end{figure}

First thought Delayed Choice experiment was proposed by Wheeler \cite{WheelersDelayedChoice} and its slightly modified version is on figure \ref{Wheelers}.

The reflected beam from the beam-splitter has changed polarization, so we put we put there half-wave plate to change it back, otherwise it would not interfere because beams with perpendicular polarizations cannot interfere. This is a special case of in principle distinguishable histories, which cannot interfere. We will talk about this experimental fact later in quantum eraser experiments.

\begin{figure}[t]
\begin{center}
\includegraphics[width=0.45\textwidth]{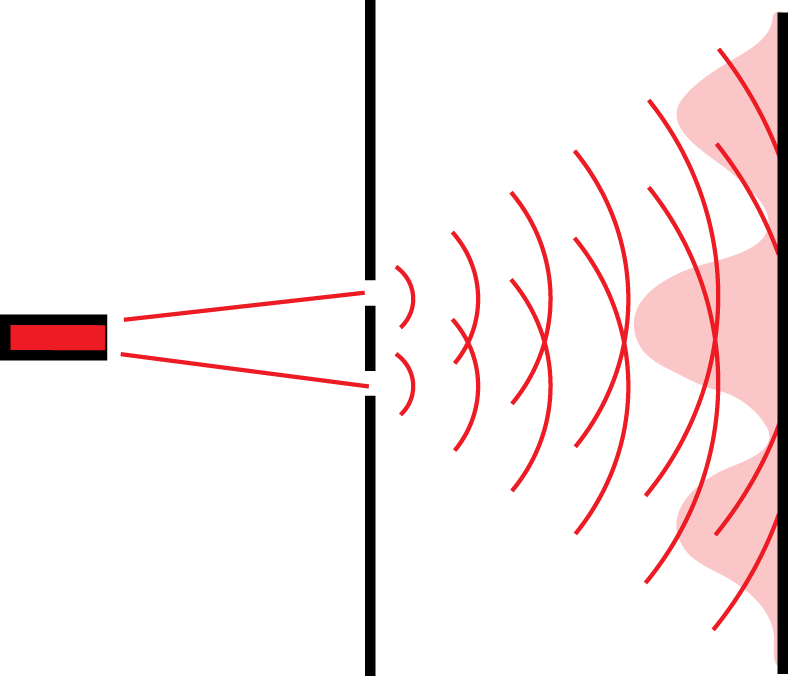}
\hspace{0.25cm}
\includegraphics[width=0.45\textwidth]{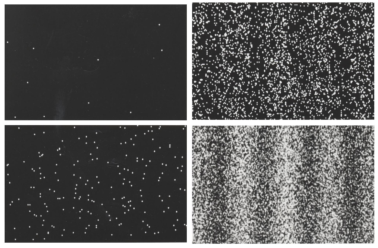}
\caption{Experimental setup of the double slit experiment and results of Akira Tonomura's realization using electrons as quantum particles. Quantum particles are red lighted.}\label{doubleslit}
\end{center}
\end{figure}

For simplicity let us consider only one photon goes through the optical system at a time. It leaves the pump beam and arrives to the beam splitter. With the classical (Newtonian) approach there is fifty percent chance of the particle going through and fifty percent chance of the particle to be reflected. So within this approach the photon goes either through the upper path or through the lower part. Then the particle arrives at lens and is focused on a rotatable mirror. With the correct rotation of the mirror we can decide which measurement we make. For example if the mirror is vertically positioned, we measure an interference pattern - which defies the Newtonian approach -- the particle must have gone through both paths at once, otherwise it could not form the interference pattern. We say that the particle behaves as a wave here. Also do not forget we had only one particle in the system, so in fact the particle hit the interference plate somewhere with certain probability and the interference pattern arise only when many such experiments are done. To illustrate that let us look to the figure \ref{doubleslit} from the double slit experiment realization with electrons.

When the mirror is horizontally positioned, we do not measure the interference pattern but rather which path the photon went. So in this experiment we decide which measurement we want to do, either interference measurement or path measurement, by rotating the mirror. Here comes the delayed choice. We can rotate the mirror not only before the photon has been released, but even after photon passed through beam-splitter and before it reached the lens. So by rotating the mirror we decide whether the photon went through the one path only (by choosing the path measurement) or went through the both paths at once (by choosing the interference measurement). But we did that after the ``decision'' of particle whether to behave like a particle or like a wave has been already done, since it already passed through the beam-splitter.

Now we have caused a lot of confusion for sure. It is because in the classical wave-particle duality interpretation one presumes it is the experimental setup which decides whether the particle is going to behave like a wave (interfere) or like a particle (going through one path only). But this delayed choice experiment shows that such prediction must be false, or at least misleading the right intuition (How could particle go through both paths at once and only through one path simultaneously?)

Now we reach the conclusion. It is not the measurement which causes the particle to behave like a wave or like a particle. It is only our interpretation we give to the particle's behavior. But in fact the particle always behave like a way (going through every possible history with certain probability), but when we look at it, we only see a particle. We see one hit at the interference plate or one hit at the detector. In other words, particles always behave like a wave when we are not looking, but when we look at them, we see only particles. The interference pattern arises only when more such experiments with more particles are done. Only the statistics of large number of particles gives us the information about the wave-like behavior. Moreover, in this experiment there is no really qualitative difference between which-path measurement and the interference measurement. Both types of detectors measure the position of the particles. The only difference is that in which-path measurement we have two detectors spatially separated. Still the saying ``The photon hit the detector D1 and thus it must have gone through lower path'' is false. All we can say is that most of the probability of hitting detector D1 comes from the lower path, but it could go through upper too. Whether the particle hits the detector D1 or detector D2 is not decided when particle goes through the system. It is decided at the place and the time of the measurement. This has also multiple philosophical consequences we will deal with in the following chapters, namely in quantum eraser experiment and relative realities approach.

\section{Quantum Eraser}

As said in the introduction, Quantum Eraser experiments are the experiments where we have some interfering system, we cancel the interference with various ways and then with certain actions on the system we may gain the interference again. We will present three examples here. Other erasers including some kind of delayed choice will be examined in the next chapter. We will see that the first two examples are not very satisfying in some sense, especially because we loose some information through the interference regaining process and one may wonder whether the regain is not just a consequence of throwing out this information. As in most articles \cite{ComplementarityHerzog, MultiparticleInterferometryGreenbergerZeilinger, KimDelayedChoiceQuantumEraser} is used the experimental fact ``only indistinguishable histories can interfere'' we will use it here too, but soon we will move to less vague and more theoretical explanation.

\subsection{Double-slit Eraser}\label{doublesliteraserSection}

\begin{figure}[!ht]
\begin{center}
\includegraphics[width=1\textwidth]{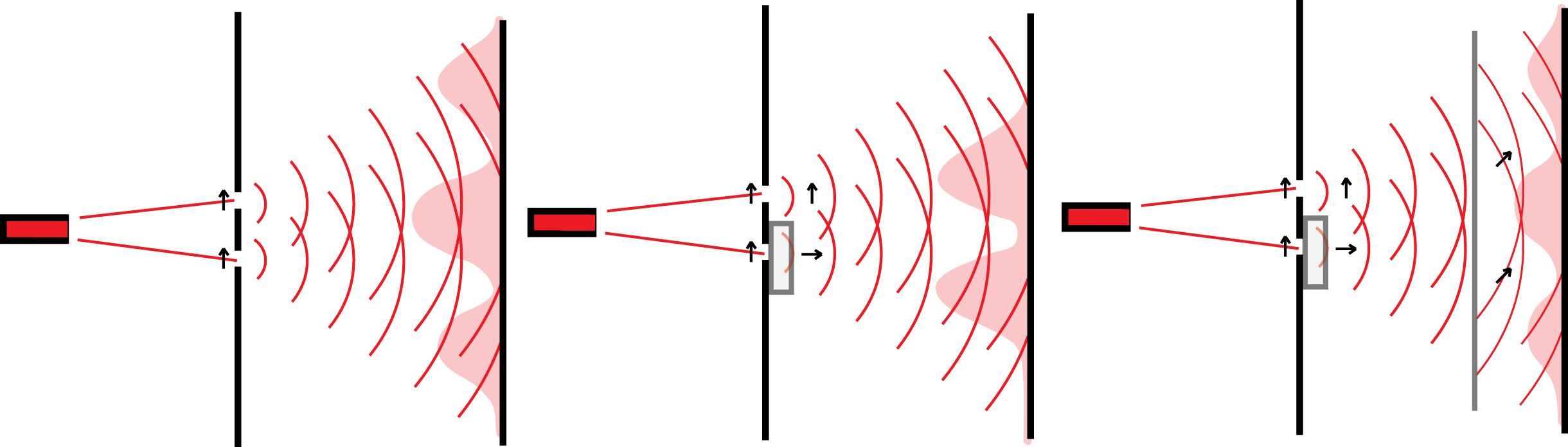}
\caption{double-slit experiment, double-slit experiment where we marked the second path using half-wave plate, double-slit with erasure of which-path information using polarization filter}\label{eraser1}
\end{center}
\end{figure}

First example is a very simple one -- just a slightly modified double-slit experiment depicted on a figure \ref{eraser1}.

For simplicity suppose the pump beam produces linearly polarized light with spin ``up''. The two paths -- going through upper slit or lower slit, are now in principle indistinguishable and thus interfere.\footnote{Do not forget it works in a very same way when only one photon is present.} But if we use quantum marker, half-wave plate, to turn the polarization of light by $90^{\circ}$, the two paths become in principle distinguishable and thus the interference pattern disappear. Nevertheless, when we erase the which-path information from the system by using polarization filter, the interference reappear.

\begin{figure}[!ht]
\begin{center}
\includegraphics[width=1\textwidth]{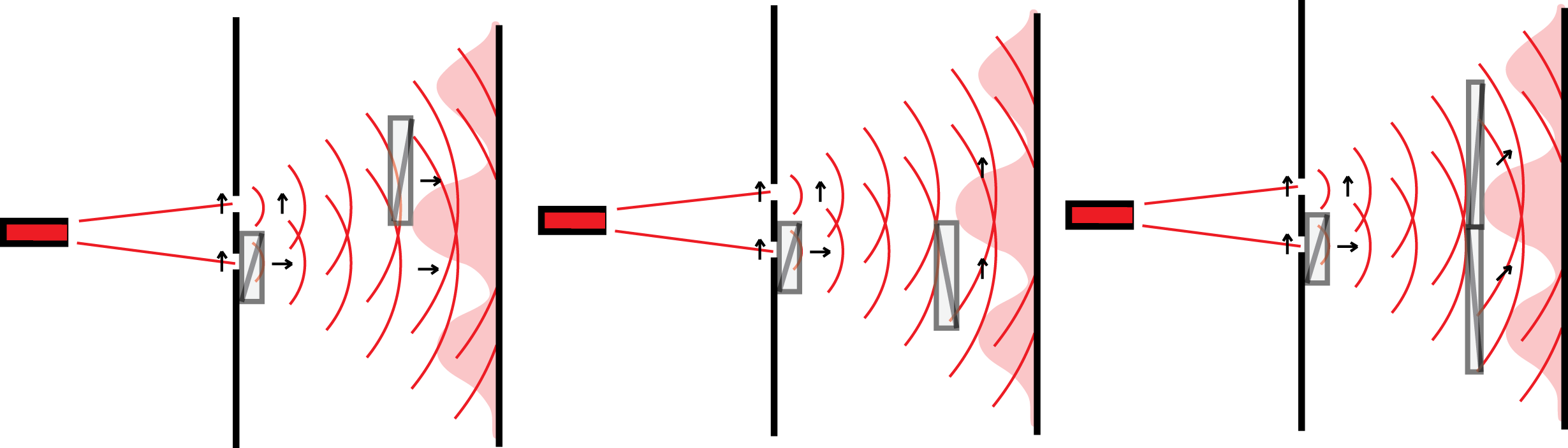}
\caption{Alternative erasures. Additional half-wave plate on the upper path, lower path and two connected quarter-wave plates.}\label{eraser1c}
\end{center}
\end{figure}

Still, this experiment is not very satisfying. It is because the polarization filter actually filters out half of the incident photons. One may ask whether the reappearance of the interference pattern is not just a simple consequence of filtering. To correct this problem we may use different instrument instead of the polarization filter. We use the half-wave plate to turn the polarization again, or we can use two connected quarter-wave plates, which in fact must be differently oriented. Alternative erasures are on figure \ref{eraser1c}.

Why we could not find something similar to polarization filter which acts in the same way to the both paths? Before we answer that question, we have to explain ``distinguishable histories'' in a proper way. ``Distinguishable histories'' means that the the which-path information about these histories are also included in some other states, which are orthogonal to each other but itself do not contribute on the actual interference. For example in the double slit eraser experiment we just presented we have interference in position. The interference here is arranged by two interfering orthogonal states $\ket{\psi_{u}}_0$ and $\ket{\psi_{d}}_0$ describing particle ``being in the upper slit'' (up) and ``being in the lower slit'' (down) respectively. These states evolves through some unitary and thus remain orthogonal as we will see later. Still they interfere. Let us denote $\ket{\psi_{u}}$ and $\ket{\psi_{d}}$ the evolved states, $\ket{\psi}=\ket{\psi_{u}}+\ket{\psi_{d}}$ wave function describing a particle going through a double slit at the time of hitting the interference plate. Intensity of point $\ket{x}$ on the interference plate is then proportional to
\begin{equation*}
\begin{split}
|\braket{x}{\psi}|^2&=|\braket{x}{\psi_u}+\braket{x}{\psi_d}|^2\\
&=|\braket{x}{\psi_u}|^2+\braket{\psi_u}{x}\braket{x}{\psi_d}+\braket{\psi_d}{x}\braket{x}{\psi_u}+|\braket{x}{\psi_d}|^2\\
&=|\braket{x}{\psi_u}|^2+2\mathrm{Re}(\braket{\psi_u}{x}\braket{x}{\psi_d})+|\braket{x}{\psi_d}|^2\ ,
\end{split}
\end{equation*}
where we see the interference term $2\mathrm{Re}(\braket{\psi_u}{x}\braket{x}{\psi_d})$.

However, in our case wave function $\ket{\psi}$ does not describe everything about our photon. Photon also has some polarization states, so the total wave function is
\[
\ket{\tilde{\psi}}_1=\ket{\uparrow}\ket{\psi}\ ,
\]
if we put nothing in front of the slits (first picture on figure \ref{eraser1}) and
\[
\ket{\tilde{\psi}}_2=\ket{\uparrow}\ket{\psi_{u}}+\ket{\rightarrow}\ket{\psi_{d}} ,
\]
if we put the half wave plate in front of the lower slit (second picture on figure \ref{eraser1}). If we take state $\ket{\tilde{\psi}}_1$ and calculate intensity of point $\ket{x}$ on the interference plate we get the same result as in the previous case without polarization. However intensity of point $\ket{x}$ with $\ket{\tilde{\psi}}_2$ is proportional to
\begin{equation*}
\begin{split}
|\braket{x}{\tilde{\psi}}_2|^2&=|\ket{\uparrow}\braket{x}{\psi_u}+\ket{\rightarrow}\braket{1}{\psi_d}|^2\\
&=\left(\braket{\psi_d}{x}\bra{\rightarrow}+\braket{\psi_u}{x}\bra{\uparrow}\right)\left(\ket{\uparrow}\braket{x}{\psi_u}+\ket{\rightarrow}\braket{x}{\psi_d}\right)\\
&=1|\braket{x}{\psi_u}|^2+0+0+1|\braket{x}{\psi_d}|^2\\
&=|\braket{x}{\psi_u}|^2+|\braket{x}{\psi_d}|^2\
\end{split}
\end{equation*}
and the interference term disappeared.

Notice that although the interfering wave functions $\ket{\psi_{u}}$, $\ket{\psi_d}$ were orthogonal, it was not them which destroyed the interference pattern. It were orthogonal states $\ket{\uparrow}$, $\ket{\rightarrow}$ which yielded the information about interfering states but did not directly take part in the interference itself (in calculations there was nothing like $\braket{x}{\uparrow}$).

In the third picture on figure \ref{eraser1} the interference reappear. It is because polarization filter acts on $\ket{\tilde{\psi}}_2$ as projector
\[
P=\frac{1}{2}\left(\ket{\bra{\rightarrow}+\bra{\uparrow}}\right)\left(\ket{\uparrow}+\ket{\rightarrow}\right)\ .
\]
As we said before, the problem of this is it filters out half of the incident photons, which also appears in a fact the resulting state $P\ket{\tilde{\psi}}_2$ is not normalized anymore.

Again, the two paths on the second picture on figure \ref{eraser1} do not interfere because spin ``up'' $\ket{\uparrow}$ and spin ``down'' $\ket{\rightarrow}$ are orthogonal to each other. It also means that we can have ``slightly distinguishable'',  ``modestly distinguishable'' and ``very distinguishable'' histories, which always refers to the percentage of non-orthogonality.\footnote{For example $\ket{0},\ket{1}$ are slightly distinguishable when $\braket{0}{1}=0.1$, modestly when $\braket{0}{1}=0.4$ etc.} We can also have slight, modest and total interference patterns as the percentage of non-orthogonality rises. Now we see that the terms of the ``distinguishable histories'' or ``distinguishable processes'' are not very satisfactory. Also some experimental realizations show that there is a continuous transition between wave-like (interference) and particle-like (non-interference) behavior, for example \cite{KaiserDelayedChoiceExperimental}.

Now back the the earlier posed question. Why there cannot be one type of substitute for the polarization filter and we need to use two quarter-wave plates glued-up together instead? We do not want to throw any information out from our system as polarization filter did and every such instrument must be expressed by a unitary operator as it is the only one which preserves all the information in the system.\footnote{We do not consider anti-unitary which are typically
 reserved to time reversal situations.} Let the $\ket{0}$ $\ket{1}$ be some orthogonal states (alternating $\ket{\uparrow}$, $\ket{\rightarrow}$). We want to find one unitary operator such that after acting on both (as the polarization filter did) the system starts interfere again. But
\[
\bra{0}U^{\dagger}U\ket{1}=\braket{0}{1}=0\ .
\]
So transformed states are orthogonal again and the system cannot interfere. Any instrument which preserves information in a system (does not filter out particles) and is used to regain the interference must act differently on the different parts of the system.

\subsection{BBO Eraser}\label{Herzogsection}

\begin{figure}[!ht]
\begin{center}
\includegraphics[width=1\textwidth]{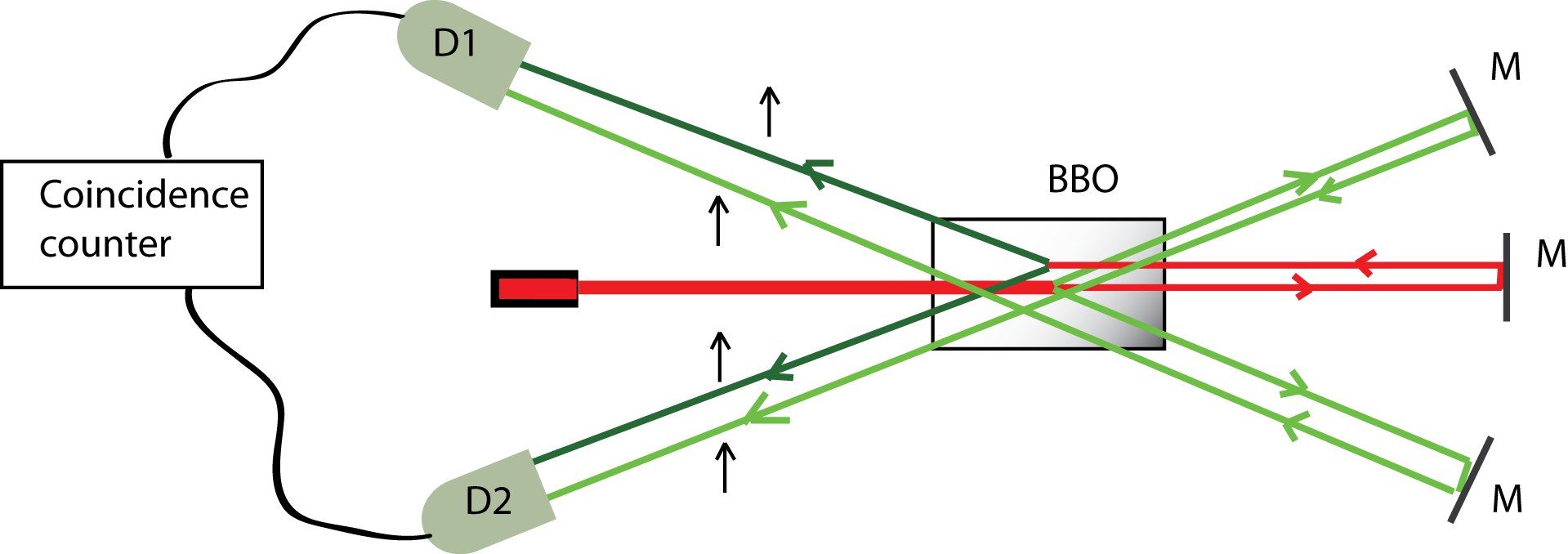}
\caption{Herzog et al. experiment without erasure. Idler interferes.}\label{Herzog1}
\end{center}
\end{figure}

The second Quantum Eraser experiment we mention here and use it later in this thesis is the experimental realization in \cite{ComplementarityHerzog}. Experimental scheme without erasure is on figure \ref{Herzog1}.

Suppose the pump beam pumps one photon at a time. This initial photon goes to the Beta-Barium Borate (BBO) crystal and has some chance to produce pair of entangled photons with correlated polarizations via parametric-down conversion. The production destroys the original photon. However, the original photon may go through crystal, be reflected by a mirror and produce the pair in a second run. So at the beginning there is one high-energy photon and at the end there are two less-energy photons. Nevertheless, in Quantum mechanics every possible history is maintained, so the pair \textbf{is} produced in the first and the second run simultaneously and these histories interfere. In other words, the photons produced in the first run (light green) of the initial high-energy photon (red) interfere with the photons produced in the second run (dark green) by the same initial high-energy photon. We will come back to this interesting fact later in the last chapter \ref{sectionRelative Realities}, which is mostly the reason why we mention this particular experiment here.

The upper photon (the one which goes to the detector D1) is usually called idler and the lower one (going to the detector D2) is called signal. Signal photon here will be used as a reference frame from which the interference on the idler can be observed. I.e. according to the article, there is $5 ns$ time window which tells us the idler and signal photon are from the same pair. We assume the lengths between BBO crystal and all of the mirrors are initially the same. However, if we move a little the lowest mirror, it causes the time interference on the idler. Idler photon can be detected before or after the signal photon is detected or even not detected within the time window at all. As we move the lowest mirror, number of detected idlers within the time window will rise and fall again finally showing the sinusoid. This is the interference as we see it.

\begin{figure}[!ht]
\begin{center}
\includegraphics[width=1\textwidth]{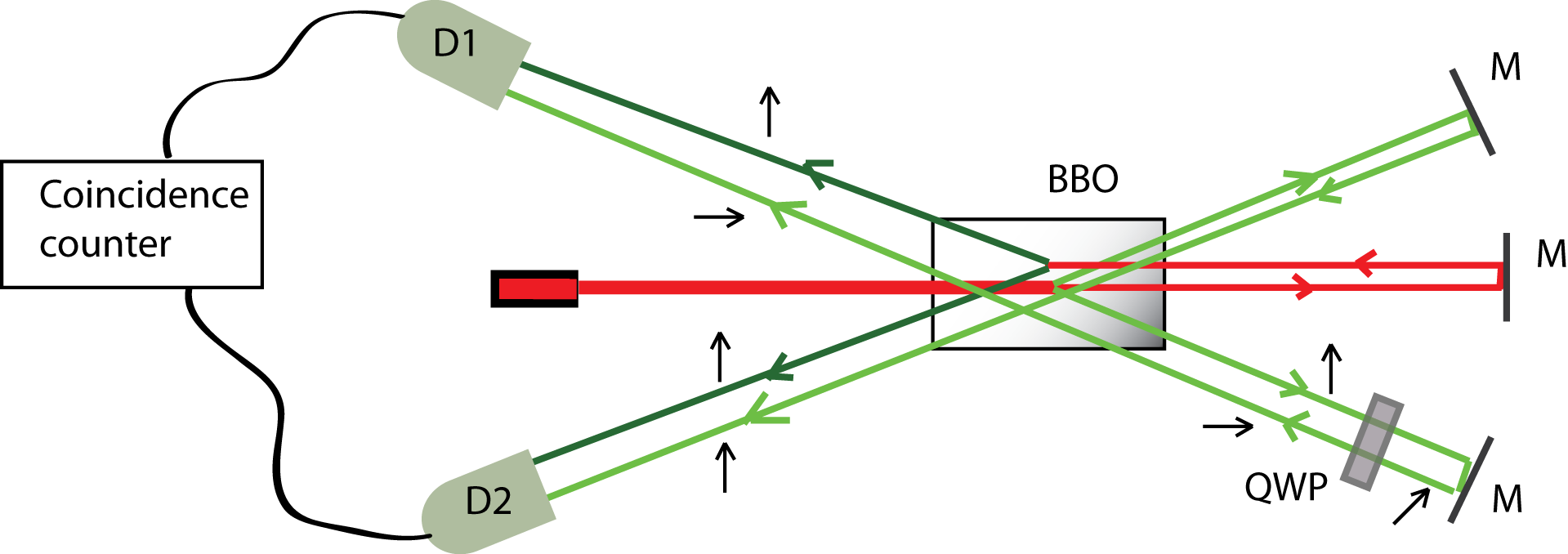}
\caption{Herzog et al. experiment with marked path. Idler does not interfere.}\label{Herzog2}
\end{center}
\end{figure}

Nevertheless, when we put the quarter-wave plate in a proper angle if front of the lowest mirror, it will turn the polarization to the perpendicular upon double passage (figure \ref{Herzog2}).

The which-path information on the idler can be in principle retrieved so it cannot interfere.

\begin{figure}[!ht]
\begin{center}
\includegraphics[width=1\textwidth]{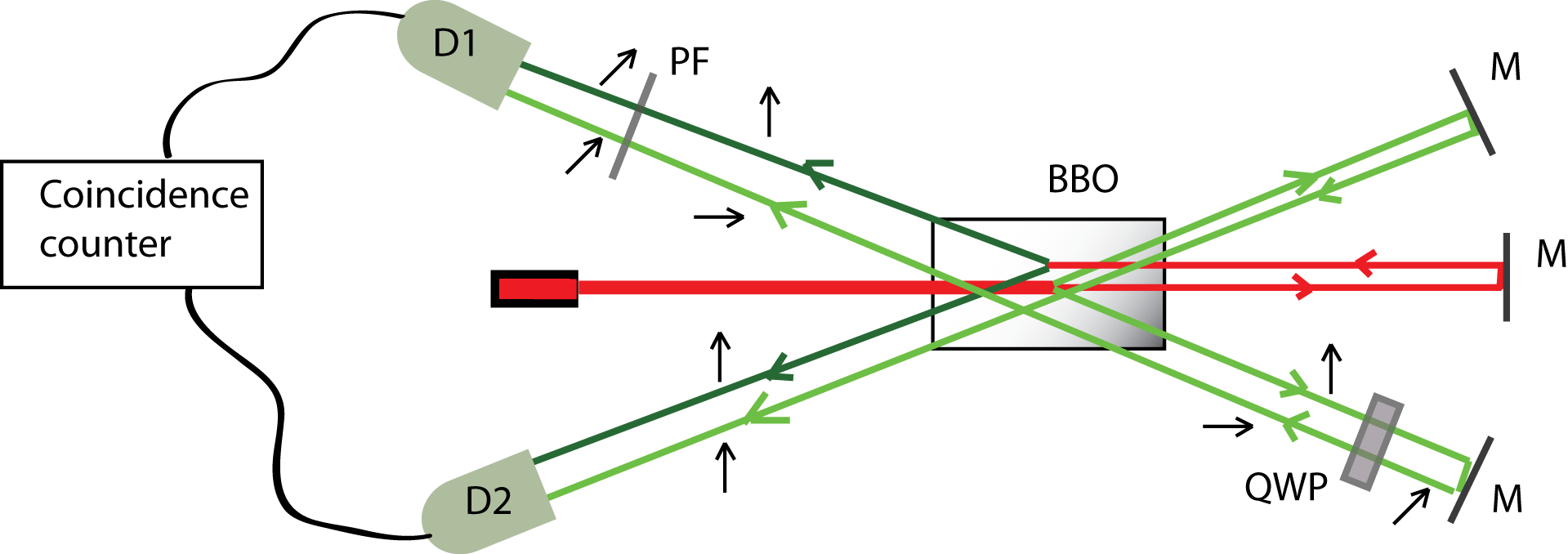}
\caption{Herzog et al. experiment, regained interference with the help of polarization filter.}\label{Herzog3}
\end{center}
\end{figure}

At last, we can erase the which-path information again using the same method we have used in the previous example. We put the polarization filter on the idler path (at 45°) and regain the interference again (figure \ref{Herzog3}).

Again, we can understand what happens without any which-path information reasoning, using just a simple mathematical analysis. Now we write all of the final states. Let $\ket{\psi}_{i,1}$, $\ket{\psi}_{i,2}$ be a wave function of the idler photon produced upon the first passage through BBO crystal, second passage respectively. Same with the signal photon $\ket{\psi}_{s,1}$, $\ket{\psi}_{s,2}$. Since the signal photon is the reference one, we can presume $\ket{\psi}_{s,1}=\ket{\psi}_{s,2}\equiv\ket{\psi}_s$.

Final state from the first experimental setup is
\[
\ket{\psi}_{i,1}\ket{\uparrow}_i\ket{\psi}_s\ket{\uparrow}_s+\ket{\psi}_{i,2}\ket{\uparrow}_i\ket{\psi}_s\ket{\uparrow}_s=
(\ket{\psi}_{i,1}+\ket{\psi}_{i,2})\ket{\uparrow}_i\ket{\psi}_s\ket{\uparrow}_s\ .
\]
The joint wave function can be written as a multiple of idler position wave-function and the rest and when we calculate probabilities in such states, interference element appears. Idler interferes.

We get the final state from the second experimental setup by a slight modification of the previous.
\[
\ket{\psi}_{i,1}\ket{\uparrow}_i\ket{\psi}_s\ket{\uparrow}_s+\ket{\psi}_{i,2}\ket{\rightarrow}_i\ket{\psi}_s\ket{\uparrow}_s
\]
This cannot be written in a form of multiplication, or, in other words, interference elements disappear as they are equal to zero due to the $\braket{\uparrow}{\rightarrow}=0$. Idler does not interfere.

We get the final state of the third setup simply from the previous just by applying polarization filter projector
\[
P=\frac{1}{2}(\bra{\uparrow}+\bra{\rightarrow})(\ket{\rightarrow}+\ket{\uparrow})_i\ .
\]
The final state is then
\[
(\ket{\psi}_{i,1}+\ket{\psi}_{i,2})(\ket{\rightarrow}+\ket{\uparrow})_i\ket{\psi}_s\ket{\uparrow}_s
\]
and the interference pattern reappear.

Note that this eraser experiment yields the same deficiency as the previous. By putting polarization filter there we actually throw out half of the initial photons.

\section{Delayed Choice Quantum Eraser experi-
\newline
-ments, Free will test}

One may wonder, could the delayed choice be used to predicting the future? Could we use such experiments to determined what the person is going to do? When the experimenter's decision affects the behavior of the particle, could not we just look at the particles behavior and find out what the experimenter is going to do? Could consequence precede the cause? Recent thought experiments which could possibly achieve that were proposed \cite{AharonovCanaFutureChoiceAffect},\cite{ZeilingerDelayedChoiceExperimentalSwapping}, nevertheless no violation of causality has been discovered. It always seem there is some tiny little thing which forbids that, still it may be pretty hard to find it. Here we will present our own thought experiment -- an attempt to beat the causality and again we will show that in this particular experiment it is not possible. However, in the next chapter we present the proof based on the Many-world interpretation which tells us all attempts to violate the causality and predict the future are doomed to failure.

\subsection{Free Will test}
\begin{figure}[h]
\begin{center}
\includegraphics[width=0.8\textwidth]{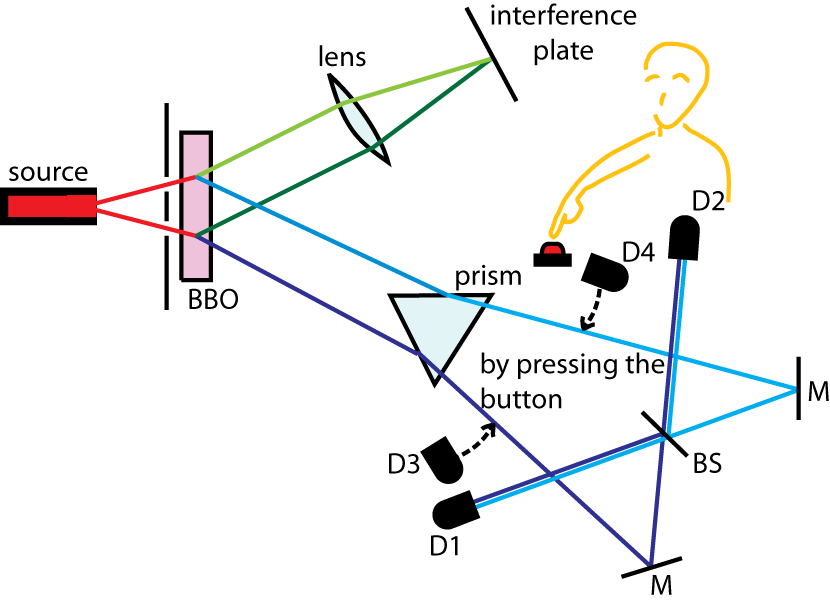}
\caption{Free Will experiment. The tested person decides whether to push the button or not. M-mirror, BS-beam-splitter.}\label{FreeWill}
\end{center}
\end{figure}

Our experimental was inspired by delayed choice quantum eraser experiment from \cite{DemystifyingTheDelayedChoice}. The scheme is on figure \ref{FreeWill}.

First to the delayed choice quantum eraser. The photon is produced in a beam splitter and goes through the double slit. We already know from the previous section \ref{sectiondelayedchoice} that both these histories are real, in other words, the photon actually goes through both slits simultaneously. Even more interesting although not so surprising is a fact that this superposition is hereditary. After the initial photon hits the BBO crystal, it produces a pair of entangled photons of which one goes up and one goes down. The initial photon went through both of the slits so the production of pair happened on two places too. Thus the produced photon is in the Bell's (maximally entangled) state
\[\label{freewill1}
\ket{UP1}\ket{UP2}+\ket{DOWN1}\ket{DOWN2}\ .
\]
We presume all of the states for each particle have the same energy so the time evolution is just an overall phase factor we can suppress (viz Appendix \ref{timeEvolution}). UP state refers to the higher place of the pair production, DOWN to the lower, 1 refers to the first photon of a pair, 2 to the second.

The idea of the free will experiment is the following. If the experimenter decides to measure the which-path information of the second photon (he pushes the red button which moves the detectors D3 and D4 into the way of approaching photon), the place of the creation of pair will be thus determined. For example if he detects the photon at the detector D4, the pair must have been created at the higher place (UP) and the first photon cannot interfere, since there is only one possible history. Put mathematically, detection of the second is equivalent of projecting the state \ref{freewill1} to
\begin{multline}
\ket{UP2}\bra{UP2}\left(\ket{UP1}\ket{UP2}+\ket{DOWN1}\ket{DOWN2}\right)\\
=\ket{UP1}\ket{UP2}\ .
\end{multline}

However, if the experimenter decides not to push the button the which-path information is erased and the first photon can interfere.

The interesting fact is we can make the path of the second photon much longer than the path of the first so we make sure the decision of the experimenter -- our tested person -- may come after the first photon hit the detector. There is the delayed choice.\footnote{In our experiment we assume energies of all photons are the same, so the evolution yields only a physically irrelevant overall phase factor.}

So could we in principle look at the interference plate and see what the tested person is going to do? When we see the interference pattern, the tested person will decide not to push the button, when we do not see it, he will decide to push it. So we could in principle determine when the tested person is going to do before he actually does it.

The problem is, it does not work, because we will never see the interference pattern on the interference plate. Let us first analyze the case where the person decides to push the button.

\begin{figure}[h]
\begin{center}
\includegraphics[width=0.9\textwidth]{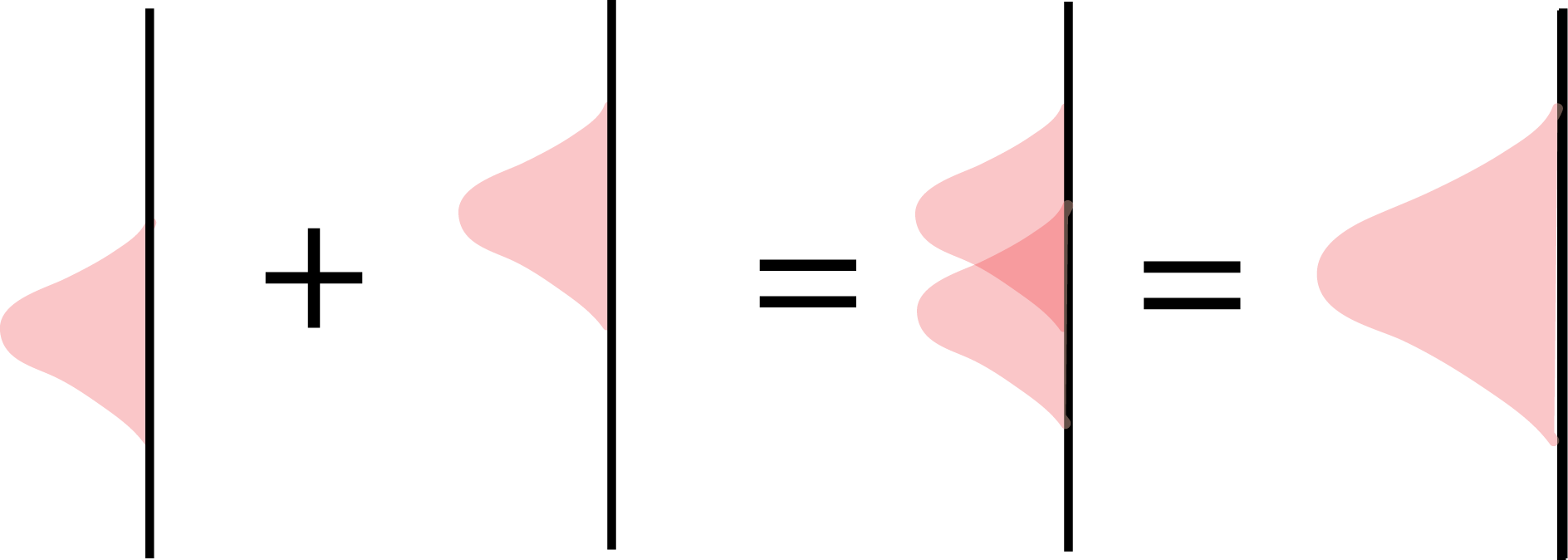}
\caption{Intensity patterns of the first photon when detector D3 clicks, D4 clicks and added together.}\label{freewillint1}
\end{center}
\end{figure}

In 50 percent of cases it is the detector D3 which makes a click. In this case we know the pair was created at the lower part and the first photon also hits the lower part of the interference plate. In 50 percent cases D4 detects a particle. The resulting intensity in these cases and their sum are depicted on figure \ref{freewillint1}.

\begin{figure}[h]
\begin{center}
\includegraphics[width=0.9\textwidth]{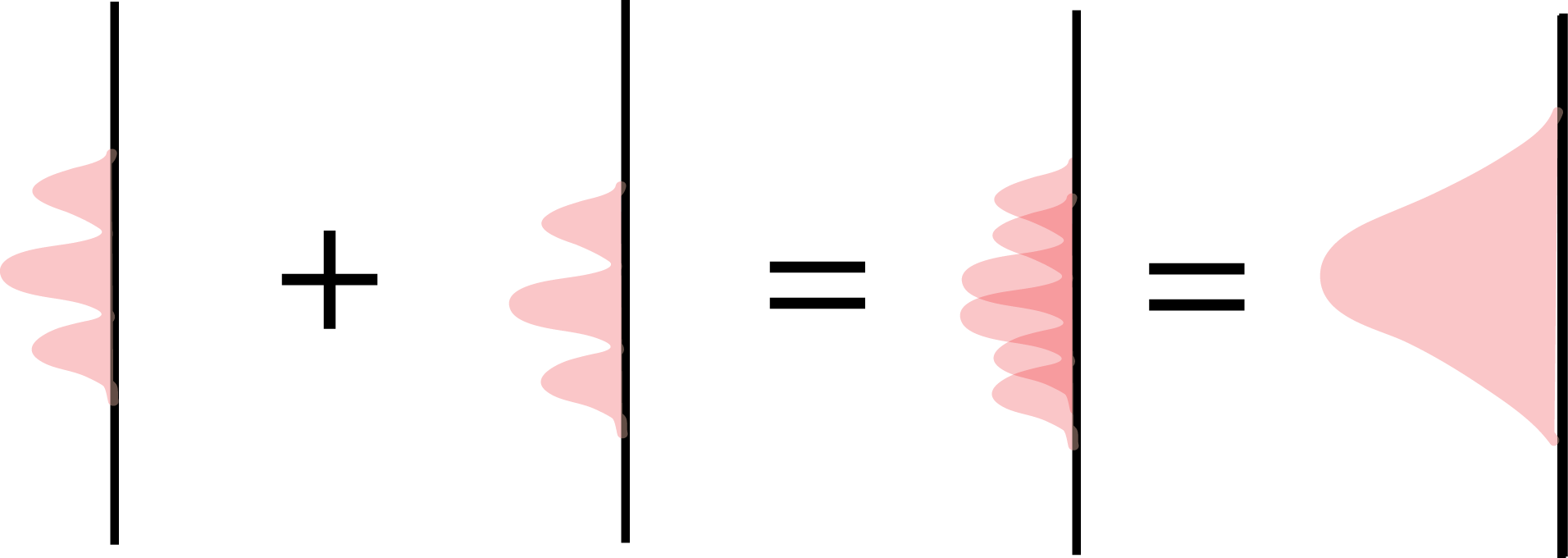}
\caption{Intensity patterns of the first photon when detector D1 clicks, D2 clicks and added together they form the very same pattern as in the previous case.}\label{freewillint2}
\end{center}
\end{figure}

If the tested person decides not to push the button, the which path information is erased and the resulting state is
\[
\ket{D1}(\ket{UP1}+\ket{DOWN1})+\ket{D2}(\ket{UP1}-\ket{DOWN1})\ .
\]
where the $\ket{D1}$ and $\ket{D2}$ are the states refering to the detection of the second photon at detector D1, D2 respectively. So in fact the first photon interferes no matter which of the detectors D1 and D2 clicks. But if the detector D1 clicks the first photon interferes differently compared to the case when D2 receive the particle. In fact, the interfering patterns are complementary and added together they form the very same shape as in the non-interfering case (figure \ref{freewillint2}).

Now let us focus a little on the conception of Delayed Choice in this experiment. Are we saying that the decision in future (push or not to push the button?) affected the result of the experiment in the past (first electron hit in a way that it forms either non-interference pattern or interference pattern)? No, this is very wrong. First of all, the we can say nothing about the decision of the tested person just from the viewing the result on the first photon. We could only say something when we already have the results measured by a second person. After we receive this result, we could retrospectively filter out our measurements on the first photon and the pattern appears. For example if the tested person just gives us list of his results in a form of zeros and ones, but he does not tell us whether he did not pushed the button or kept it pushed, we could say what he did just from filtering half of our results. We can throw out all of the results from pairs where he received zero. Then if we see the interference pattern, we can be sure he did not push the button. If we see just one peak we can be sure he kept the button pushed. But this could happen after we received his results and as we know, he could not give us his result faster than the speed of light. So no prediction of future is happening. Because to predict what will happen we need to know first something from this future.

Other reason why usual interpretation of Delayed Choice experiments as the experiments where future results affect the past fails is because the case is totaly symmetrical. For instance suppose the tested person decided not to push the button. We say that if the detector D1 clicked (in future) then we have quite a big chance that the first photon hit somewhere (in past) where the intensity depicted on the very left picture of figure \ref{freewillint2} is the highest. But we also turn this over so we do not get any strange interpretation. We can say that if the first photon hit somewhere where the intensity of the very left picture is the highest, we have quite a big chance that the second photon will be detected at D1. And this case is not paradoxical at all.

Entanglement is symmetric and somehow out of time. It does not matter which particle from the pair do we measure first. It is just our interpretation what changes, but not real measurable quantities. The entangled nature comes out only when both experimenters controlling parts of the entangled pair come together and compare their results. From measurement of one particle from pair we cannot say what someone decides to do to the second. We only can predict what the this person will measure, but only if we know in advance what he is going to do.

The entangled nature is somehow different from the real world. As EPR paradox and violating Bell's inequalities \cite{PalssonViolatingBellExperimental} shows there must be some hidden interaction between the pair, obviously faster than light, but we cannot use this interaction for any causality violations nor for predicting the future. We will come back to this in the next chapter and generalize this from one example showed here to any entangled system using the Many World interpretation and presumption of unitary evolution of the closed system \ref{sectionNoCommunication}.

In the next section we will present the last delayed choice quantum eraser we will deal with here. It will be another example and maybe even better to demonstrate the ideas gained here.

\subsection{Delayed Choice entanglement swapping}

The basic idea of delayed choice entanglement swapping is to swap the entanglement \emph{after} one of the particle of entangled pair has been already measured and thus the entanglement may not\footnote{in the classical point of view where the measurement leads to the instantaneous collapse of the wave function} longer exist. The original though experiment was introduced by Peres \cite{PeresDelayedChoiceEntanglementSwapping} but we will utilize mainly the experimental realization presented in \cite{ZeilingerDelayedChoiceExperimentalSwapping}.

\begin{figure}[h]
\begin{center}
\includegraphics[width=0.9\textwidth]{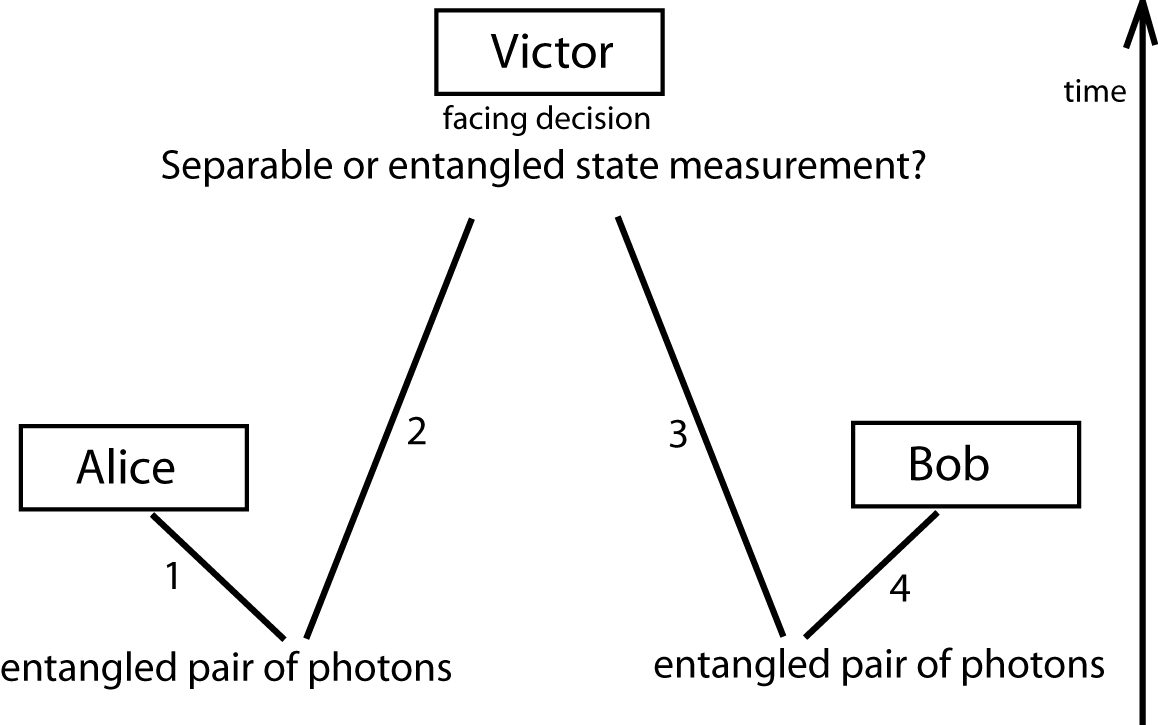}
\caption{Delayed choice entanglement swapping.}\label{entanglementswapping}
\end{center}
\end{figure}

The basic scheme is on figure \ref{entanglementswapping}

In this experiment we produce two pairs of entangled photons and Alice measures polarization of the first and Bob of the fourth. Second and third photon (one of each pair) go to Victor and Victor decides whether to subject them to entangled-state measurement (projecting them onto one of the Bell's states) or separate state measurement (projecting them tensor product of the polarizations of the second and third). Note that this is the very analog of the decision of our tested person in Free will experiment.

Now we present confusing explanation what is happening to clarify it later. If Victor decides to do the entangled-state measurement, the first and fourth will entangle, even they have not never met. And according to the experiment setup, Victor can entangle them even after the polarization of the first and the fourth has been actually measured!

To clarify, let us do the same reasoning we did in the Free Will experiment. First of all, if the Alice and Bob would be able to say whether their photons are entangled or not, they would be able to predict the (future) Victor's decision. From our former experience we assume that this is not possible. Indeed, the analysis of this experiment shows that they cannot be able to predict what will Victor do. The analysis is quite straightforward but very long so we will not deal with it here. Actually, to be able to say the first and fourth photon are entangled we need to have Victor's measurement results first. After we have it, we can sort the Alice's and Bob's measurement into two groups -- group where Victor measured zeros and group where Victor measured ones. After we have these two groups we can say whether the first and fourth photon are entangled or not. Without much surprise we discover in case when Victor decided to do the entangled-state measurement the first and fourth photon are correlated (and thus must have been entangled) and where Victor decided to do the separable-state measurement there is not correlation at all.

So did we discover that future decision affected the entanglement of the already destroyed (measured) states? Sort of. Still such statement is very misleading since there is not causality issue here again (we need Victor's results first to say whether or not they are entangled) and again, as in the Free Will experiment, the affecting is symmetrical. We can also say that the results of the experiments Alice and Bob did on their first and fourth photon affected the result Victor received. And this is true indeed. In confusing explanation we say the Victor's decision affected the entanglement of the Alice's and Bob's photon but to prove that we need to have Victor's measurement results first. But these results were affected by the results of Alice's and Bob's measurements! It is the same to say the consequence affected the cause. In fact, It does not matter what we call the cause and what the consequence since there is not any causal relation. Only our interpretation is sometimes confusing, because without much success we are trying to replace correlations and causal influence which are in its essence unmistakable.


\chapter{General view on Delayed Choice and Quantum eraser experiments}

In the previous chapter we presented many Delayed Choice experiments and we showed that in these particular experiments causality is not violated. We cannot use them to predict the future, we cannot use them to send information back in time, we cannot use them to determine whether the tested person has or has not the free will. But is it a general rule that entanglement cannot be used for such wonderful purposes? Yes it is. At least when you accept the presumption of unitary evolution of the closed system and the concept that measurement is just an interaction between quantum system and the observer, which is quantum system itself.

\section{Why predicting the future does not work? Answer: No-communication Theorem}\label{sectionNoCommunication}

\begin{figure}[h]
\begin{center}
\includegraphics[width=0.9\textwidth]{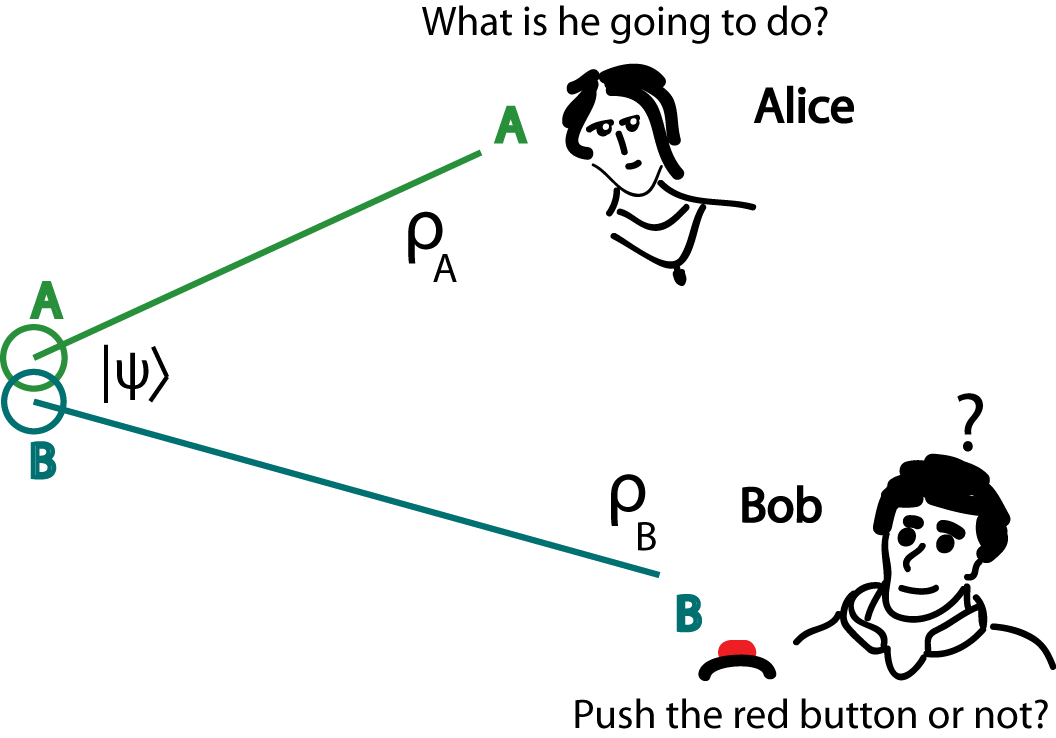}
\caption{Generalized Scheme of Delayed Choice experiments used for predicting the future using entanglement.}\label{generalizedscheme}
\end{center}
\end{figure}

There were two kinds of Delayed Choice experiments, first type, as the Wheeler's delayed choice, included only one particle and as such it cannot be used to predict the future in any manner. However, the second type using entangled particles could be in principle used for prediction. We already presented two such experiments - the Free will experiment and Delayed Choice Entanglement Swapping. However in both of these we showed we cannot beat the causality. We will generalize this to any system using the entanglement. To do that, let us present the generalized scheme on figure \ref{generalizedscheme}.

At the beginning there is some process producing an entangled system of two or more particles. It does not matter how the two systems $A$ and $B$ became entangled, nor what kind of the entanglement we deal with here. We suppose the joint state of systems $A$ and $B$ is pure and is described by a wave function $\ket{\psi}$. If Alice could observe the wave function itself, she would be able to notice the instantaneous collapse of it when the Bob measures his part, even if the measurement was done in future. That itself is an interesting fact so we will highlight it here.

\bigskip
\textbf{Direct observing of a wave function would lead to predicting the future.}
\bigskip

Nevertheless, she is not able to see it. All she can see are some hits or any other types of measurement results and these results are somewhat random, although following some pattern. We will assume that at the time of the measuring the systems $A$ and $B$ are no longer interacting. If this were not true, some information from the Alice's system could reach the Bob's and affect his decision and we want to let him decide on his own. Let us derive whether Alice will be able to determine from her results what Bob is going to do.

The whole system is in a pure state $\ket{\psi}$ but Alice has only access to a part of it, to the system $A$. The whole system is thus described by a density operator
\[
\rho=\ket{\psi}\bra{\psi}\ ,
\]
and her system before the measurement is described by this density operator traced over the system $B$
\[
\rho_A=\mathrm{Tr}_B(\rho)\ .
\]
All probabilities of her results are governed by this density operator, so the question is: Will she be able to notice the change when Bob chooses to push the red button? In other words, will the density operator change?

Let us first suppose Bob decides whether to apply on his part $B$ of the system some unitary operator $U$ or not to apply anything. The unitary operator will act on the whole system as a tensor product $\tilde{U}=\mathbf{1}\otimes U$. Alice's density operator after his decision to apply is

\begin{equation}\label{doesnotchange}
\rho_{A\text{ after application}}=\mathrm{Tr}_B(\tilde{U}\rho\tilde{U}^{\dagger})=\mathrm{Tr}_B(\rho\tilde{U}^{\dagger}\tilde{U})
=\mathrm{Tr}_B(\rho)
\end{equation}

We owe the second equality to the special form of $\tilde{U}$ as a tensor product, since in general a partial trace is not invariant under  cyclic permutations. We see that in this case Alice's density operator describing her system $A$ does not change so she is observers the very same results as if Bob decided to do nothing. She cannot predict anything.

Maybe you noticed we silently omitted the unitary evolution of the system itself. Let us set it right now. We assumed our system is no longer interacting, in other words, it evolves through the unitary which is the tensor product of unitary operators acting on system $A$ and system $B$. If we take the time evolution into consideration, the derivation of (\ref{doesnotchange}) will be the very same, only a bit longer because we would need to take the time into account.

Now to the generalized operation. Let us suppose Bob is going to do something non-unitary, i.e. measurement. He decides whether to do one type of measurement or the other. In Free will experiment it was the decision whether to push the button and measure the which-path information or do not push anything and erase this information. In Delayed Choice Entanglement Swapping it was the Victors decision whether to do separable-state or entangled-state measurements (and the role of Alice in our generalized scheme play Alice and Bob together). Now we use the basic presumption of the Many World interpretation. Measurement is just an interaction between observer and the system from the view of the observer. It seems non-unitary to the observer because the observer is involved in it. But from the outside the interaction between the observer and his measured system is just a unitary evolution on the joint system Observer$+$observed system. So here Bob and system $B$ together act as a closed system (not interacting with other systems, in particular system $A$) and from the point of view of Alice the joint system evolves through unitary. We can just expand the system $B$ to a new one $\tilde{B}=B+\mathrm{Bob}$ and derive the very same equality for the density operator describing the system A. Again, Alice is not able to notice any change and thus cannot determine what Bob is going to do.

The derived theorem is called no-communication theorem and is usually proven using set of Kraus operators\footnote{$\{K_1,\dots,K_m\}$ such that $K_1^{\dagger}K_1+K_2^{\dagger}K_2+\cdots+K_m^{\dagger}K_m=\mathbf{1}$} which act as the most generalized measurement one is able to do. The advantage of our proof is that we do not have to presume any form of the most generalized measurement, instead we take much simpler presumption of unitary evolution of a closed system and presumption that people/any other observer behave in the very same way as the rest since they are made of the same atoms and there is not reason why they should have behave differently.

The name of the theorem fits very well. It actually says that no information can be passed through the entanglement alone. In other words, two parties cannot communicate using just an entangled systems alone.

This theorem solves the causality questions in our Free Will experiment and in many articles too, namely \cite{ZeilingerDelayedChoiceExperimentalSwapping, KaiserDelayedChoiceExperimental, AharonovCanaFutureChoiceAffect, PeruzzoDelayedChoiceExperimental} In all of these experiments they showed there is not any causal link. Now we know if we design new experiment with similar principles, it will behave in a same way. There will be no causal link between two entangled particles and one cannot use just the entanglement to predict what the other person is going to do.

Notice one interesting thing. We presumed the unitary evolution of Bob is entirely deterministic, in other words, his unitary evolution with the system $B$ guaranteed Alice was not be able to determine his decision, she cannot predict what he is going to do, his decision is entirely non-deterministic for her. It seems the deterministic evolution of Bob gives him a chance to have a free will. Still, strictly speaking, if there was a minimal chance he decides either way, both of his decisions are made with certain probability and both are real, as in quantum mechanics every possibility exists simultaneously as proved in the first section in Wheeler's delayed choice experiment. Maybe free will is just an ability to choose the outcome of your measurement.

Let us summarize this whole section in one last sentence.

\bigskip
\textbf{Entanglement alone in Delayed Choice experiments cannot be used for prediction of the future, all such experiments trying to do that are doomed to failure.}
\bigskip

\section{Complementarity of one-particle interference and correlations}\label{sectionComplementarityofinterferenceandcorrelations}

Here we will show using the same presumption as in the previous section often used experimental fact that one-particle interference and two-particle interference (correlations between two particles) are complementary, in other words, we cannot observe the interference pattern of one particle while it is correlated with some other. More precisely speaking, we can observe it, but better visible interference pattern would lead to worse visible correlations and vice versa. Perfect interference pattern of one particle leads to no correlation with any other.\footnote{In an observable we measure. For example if we measure the polarization of a photon, it does not mean its energy is not entangled and thus correlated with some other particles.}

\begin{figure}[h]
\begin{center}
\includegraphics[width=0.9\textwidth]{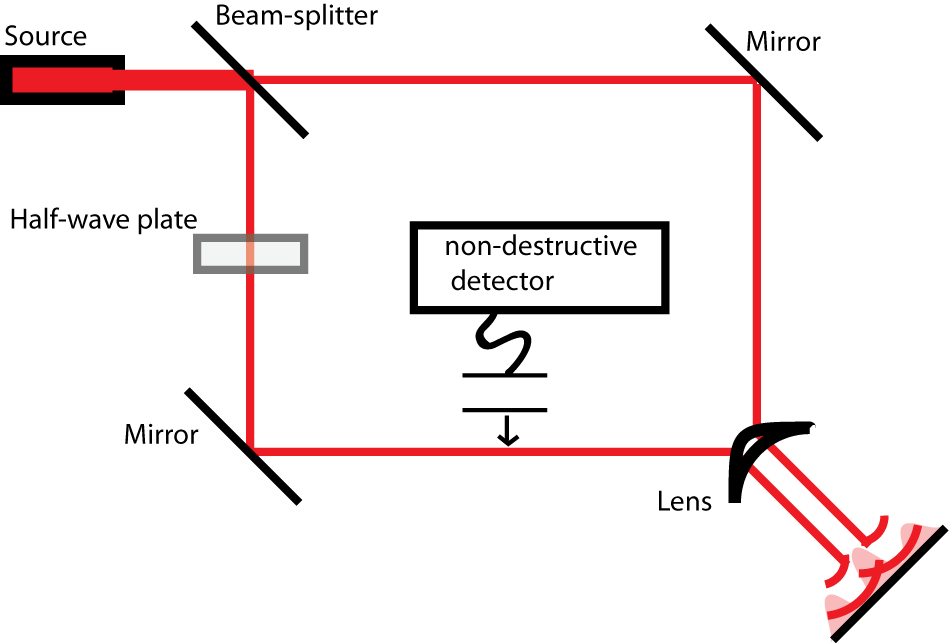}
\caption{Double-slit experiment with interfering paths.}\label{entanglementdoubleslit}
\end{center}
\end{figure}

First we demonstrate the basic ideas on the well known experiment -- the double slit experiment. We just use a slight modification of this experiment as we also did in the Wheeler's delayed choice. The scheme is on figure \ref{entanglementdoubleslit}. It is well known fact that even non-destructive measurement of the which-path information prevents the system from interfering. Here we show that one-particle and two-particle interference complementarity and the non-destructive measurement destroying the interference are basically the same thing.

On the figure we see the double-slit experiment without slits but beam-splitters instead, still the idea is identical. Let us denote $\ket{\psi_u(0)}$ the upper path of the photon and $\ket{\psi_d(0)}$ the lower path. We have also some non-destructive measurement device there which could measure in which way the photon goes.\footnote{Since photons interact only weakly one can, for the sake of argument, assume electrons instead. Our reasonings will not be influenced by this assumption.} Although if we do not put it in the photon's way, the two waves evolved from $\ket{\psi_u(0)}$ and $\ket{\psi_d(0)}$  finally overlap each other and interferes on the interference plate. So we have one-particle interference. Let us denote evolved states as $\ket{\psi_u}$, $\ket{\psi_d}$.\footnote{Now we will apply the very same derivation we used in section \ref{doublesliteraserSection} of the first chapter when we described distinguishable histories. The difference now is that instead of polarization states of our interfering particle we now use some additional particle/particles we call measurement device and we will write $\ket{0}$ instead of $\ket{\uparrow}$ and $\ket{1}$ instead of $\ket{\rightarrow}$. So the difference is in type of system we describe and different goal, math remains the same, at least in the beginning.} The state final state of a particle is then a simple sum $\ket{\psi}=\ket{\psi_u}+\ket{\psi_d}$ Probability of measuring a particle hitting the position $x$ on the interference plate is then proportional to\footnote{We say proportional because we did not normalize our states. In our consideration we, however, do not need to work with normalized states and so, for the sake of simplicity, we restrain from using the normalization.}
\begin{equation*}
\begin{split}
|\braket{x}{\psi}|^2&=|\braket{x}{\psi_u}+\braket{x}{\psi_d}|^2\\
&=|\braket{x}{\psi_u}|^2+\braket{\psi_u}{x}\braket{x}{\psi_d}+\braket{\psi_d}{x}\braket{x}{\psi_u}+|\braket{x}{\psi_d}|^2\\
&=|\braket{x}{\psi_u}|^2+2\mathrm{Re}(\braket{\psi_u}{x}\braket{x}{\psi_d})+|\braket{x}{\psi_d}|^2\ ,
\end{split}
\end{equation*}
where we can see the interference component $2\mathrm{Re}(\braket{\psi_u}{x}\braket{x}{\psi_d})$.

However, if we put detector in a way of the photon, the interference disappear. The interesting fact remains our detector needs not to be consisted of millions of particles, one or few particles totaly suffice. In the beginning of this thesis we assumed that since everything is made of the same type of particles, we can threat the detector consisted of million particles in a same way as the one particle only. This approach will be even more justified in the next chapter where we present Wigner's friend approach to the measurement. Suppose that the detector can be in two orthonormal states
$\ket{0}$,
which is its initial state and it remains if it detected nothing and
$\ket{1}$,
if it detected a particle.

These are the basis states we measure when we look at our detector and it is unimportant whether it is only few particles or macroscopic system.

So the whole system after measurement (unitary interaction \ref{unitary1} between the particle and the measurement device, which itself can be a single particle) is in a joint state
\[
\ket{\psi}=\ket{0}\ket{\psi_u}+\ket{1}\ket{\psi_d}\ .
\]
For now for simplicity we suppose the evolution of measuring device after interaction (or measuring particle if you like) is trivial, i. e. it consists only from overall phase factor we can suppress. We will come to that later.

Now our particle cannot interfere, since the probability is now proportional to
\begin{equation*}
\begin{split}
|\braket{x}{\psi}|^2&=|\ket{0}\braket{x}{\psi_u}+\ket{1}\braket{1}{\psi_d}|^2\\
&=(\braket{\psi_d}{x}\bra{1}+\braket{\psi_u}{x}\bra{0})(\ket{0}\braket{x}{\psi_u}+\ket{1}\braket{x}{\psi_d})\\
&=1|\braket{x}{\psi_u}|^2+0+0+1|\braket{x}{\psi_d}|^2\\
&=|\braket{x}{\psi_u}|^2+|\braket{x}{\psi_d}|^2\ ,
\end{split}
\end{equation*}
and the interference term disappeared. So we do not have one-particle interference in this case, but we have correlations, two-particle interference in other words. We may not be able to distinguish between the states $\ket{\psi_u}$ and $\ket{\psi_d}$ perfectly, for example in this double slit experiment the probability amplitudes may overlap as on the picture \ref{freewillint1}, still for position $\ket{x}$ sufficiently high on the interference plate we have
\[
P(0|x)=P(x|0)\approx 1.
\]

So the complementarity of the one-particle and two-particle interference is equivalent to the complementarity of interaction or non-interaction between the two particles\footnote{In our case it was it was one photon/electron and measurement device consisting of one or more particles.} or complementarity of common origin or not common origin\footnote{As the entanglement of two photon produced via parametric-down conversion in BBO crystal.}.

We must note one important fact. If we choose to measure our measurement device in some different basis, for example
\[
\ket{+}=\frac{1}{\sqrt{2}}(\ket{0}+\ket{1}),\ \ket{-}=\frac{1}{\sqrt{2}}(\ket{0}-\ket{1})\ ,
\]
the correlations will change as well. For example $\ket{+}$ will be now correlated with the $(\ket{\psi_u}+\ket{\psi_d})$, which forms one of the sinusoid depicted on figure \ref{freewillint2}. The correlations would have also appear differently if the evolution of the detector was not trivial. In which form the correlations appear depends on both unitary evolution and the basis in which we measure, but the fact remains, no unitary evolution acting only on the measurement device alone (of form $U\otimes\mathbf{1}$ on a whole system) can restore interference, since unitary evolution preserves orthogonality. You may also ask: ``Could some, not necessarily unitary evolution, but some action on the measuring device alone restore the interference on our formerly interfering particle?'' No. Because this non-unitary action still must be unitary in some bigger system. This reasoning led us to the No-communication theorem which itself answers that posed question alone. Here we just see another explicit application of that.

Now we will show on an example we can continuously move between all correlation-like behavior and all interference-behavior and also that the exact amount between of interference and correlation depends on the interaction between the two particles. In other words, we will show that better interference pattern leads to worse correlations and vice versa.

Let us suppose initial state (before the interaction)
\[
\ket{i}=N_i(\ket{0}+\ket{1})(\ket{\psi_1}+\ket{\psi_2})
\]
going onto final state (after the interaction)
\[
\ket{f}=N_f(\ket{0}\ket{\psi_1}+\ket{1}\ket{\psi_2})
\]
via some unitary transformation, where $N_i,\ N_f$ are the normalization constants,
\[
\ket{\psi_1}=\ket{L}
\]
\[
\ket{\psi_2}=\frac{3}{5}\ket{L}+\frac{4}{5}\ket{R}\ ,
\]
where $\ket{L},\ \ket{R}$ are orthonormal vectors. In the previous case $\ket{0},\ket{1}$ were two states of the detector, but now strictly speaking we cannot call this detector again, since detector should project on the orthogonal states, but $\ket{\psi_1}$ and $\ket{\psi_2}$ are not longer orthogonal here. Still, the unitary transformation between the initial and final state exists and one of the many possible is written in the appendix \ref{unitary2}. After a successful normalization we get
\[
\ket{i}=\frac{1}{\sqrt{10}}\left(\ket{0}+\ket{1}\right)\left(2\ket{L}+\ket{R}\right)
\]
\[
\ket{f}=\frac{1}{\sqrt{2}}\ket{0}\ket{L}+\frac{3}{5\sqrt{2}}\ket{1}\ket{L}+\frac{2\sqrt{2}}{5}\ket{1}\ket{R}\ .
\]

From the look of the final state we are sure that we could see some correlations. The exact amount of entanglement can be calculated using for example relative Von Neumann entropy \cite{NielsenChuang} but we will not do it here. We will rather look on a change of the visibility of the interference pattern. Let us suppose we measure observable $A$ on the second system. Probability of obtaining a value $a$ on the initial state is proportional to
\begin{equation*}
\begin{split}
|\braket{a}{i}|^2&\propto|2\braket{a}{L}+\braket{a}{R}|^2\\
&=4|\braket{a}{L}|^2+2\braket{a}{L}\braket{a}{R}+2\braket{R}{a}\braket{L}{a}+|\braket{a}{R}|^2\\
&=4|\braket{a}{L}|^2+4\mathrm{Re}(\braket{a}{L}\braket{a}{R})+|\braket{a}{R}|^2\\
&\propto\frac{4}{9}|\braket{a}{L}|^2+\frac{4}{9}\mathrm{Re}(\braket{a}{L}\braket{a}{R})+\frac{1}{9}|\braket{a}{R}|^2\ .
\end{split}
\end{equation*}

Probability of obtaining the same value in final state is proportional to
\begin{equation*}
\begin{split}
|\braket{a}{f}|^2&\propto|5\ket{0}\braket{a}{L}+3\ket{1}\braket{a}{L}+4\ket{1}\braket{a}{R}|^2\\
&=(4\braket{R}{a}\bra{1}+3\braket{L}{a}\bra{1}+5\braket{L}{a}\bra{0})\cdot\\
&~\cdot(5\ket{0}\braket{a}{L}+3\ket{1}\braket{a}{L}+4\ket{1}\braket{a}{R})\\
&=34|\braket{a}{L}|^2+24\mathrm{Re}(\braket{a}{L}\braket{a}{R})+16|\braket{a}{R}|^2\\
&\propto\frac{17}{37}|\braket{a}{L}|^2+\frac{12}{37}\mathrm{Re}(\braket{a}{L}\braket{a}{R})+\frac{8}{37}|\braket{a}{R}|^2\ .
\end{split}
\end{equation*}

Now we see that we were able to find a state which interferes as well as the results of its measurements are correlated with some other. The initial state was not correlated at all. However, to impose some correlation between the two systems we had to pay by a worse visible pattern on a second. Our trade off is
\[
\frac{12}{37}=0.32<0.44=\frac{4}{9}\ .
\]

It would be interesting to find a precise relationship between amount of correlations and amount of interference. This is the task for the future.

At the end of this section let us note that in the beginning we used the very same method the Decoherence Theory use to explain the measurement \cite{SchlosshauerDecoherenceMeasurementInterpretations}. The Decoherence Theory says measurement is a random unitary interaction of our system (particle) with the environment consisted of millions of particles. To properly describe the system we should work with a huge wave function including our system and the millions of particles, but because we are interested only in our system, we trace over the environment and acquire reduced density operator, generally mixed state. Due to the vast amount of particles in the environment is our particle in one of the orthogonal state -- one of the eigenvectors of the reduced density operator. The difference is our environment, i. e. non-destructive measurement device was not necessarily consisted of vast amount of particles, one or few particles were sufficient. That was also a reason why we were able to describe something as a correlation. Loosely speaking, information about correlation between particle and the environment consisted of millions of particles is lost in numbers. However, even though Decoherence Theory describes why the particle passes onto one of the orthogonal states after the measurement, it still does not explain why it passes onto one particular.

\chapter{Relative Realities}

In this chapter we take the knowledge gained from studying delayed choice and mainly quantum eraser experiments to introduce new concept of looking at the Quantum Mechanics. There are many theories, or rather meta-theories describing what is really happening in with their own views at the measurement problem. Namely Decoherence theory, which explains the measurement as an act of particle randomly interacting with huge amounts of other particles and from this concept one is able to derive Born projection rule, but it does not explain the measurement itself and its probability character. On the other hand other theory, well known Many World interpretation is going further (although it is older) and is saying an act of measurement is an act where one world splits in two, where each of the new worlds yields the different outcome of the measurement. The World splits infinitely to the new worlds, in which every possibility is realized. Other interpretation, Consistent Histories approach rather use Propositions as the fundamental issue in Quantum Mechanics, while some propositions together do make sense (the histories are consistent) and some do not.

There are many theories, but the goal of this chapter is not to contradict any of these theories, rather to broaden their perspective. We will be closest to Many World interpretation, still, we will show that this interpretation leads to some unsolved questions, mainly due to the lack of precise explanation what the measurement is. Many world interpretation also says the splitted worlds are no longer independent, cannot communicate with each other and once the splitting is done there is no turning back. We will disprove the last statement and rather propose another approach. We will show the turning back does not contradict anything and does not lead to any kinds of logical paradoxes and we will also have experimental support for this hypothesis. The tool of this turning will be nothing else than well-understood Quantum Eraser experiments, only applied on bigger systems, which we assume must follow the same laws as the quantum tiny world.

Since antique period people are dealing with issue of relativity. Maybe one of the first discovery was the relativity of distances. One thing may closer for one person, farther for the other. There is no doubt already cavemen used this idea. Still it took centuries when people discovered the movement is also relative. We just should not say ``The car is moving by 100 km per hour.'', because this sentence does not make sense. What we should say is ``The car is moving by 100km per hour with respect to ground.'' When we talk about cars we usually mean it in the latter sense, but we could also think about car moving with respect to Sun by $107,300$ km per hour. The proper understanding of this knowledge came after Newton in seventeenth century. Two hundred years later Einstein discovered that not only movement is relative, but also lengths and time are and some people could possibly use this knowledge to outlive their grand grand grandchildren. Due to this, we also know wave functions must be relative for different observers, since for a moving observer wave function of some wave packet is squeezed, for observers who posses some additional information the wave function is the conditional one, for example knowing outcome of the measurement on the entangled pair will allow us to describe our particle better, while others are forced to use density operator to describe the same particle. We will show is the reality is also relative. For one observer there may exists a particle which do not for other, and the same particle may exist for you in one time, but do not exist in other. But to do these tricks we are forced to follow the rules which guarantee no logical paradox can occur. The main conclusion of this chapter is there is not an absolute reality as there is not an absolute time. We always need to relate the reality to a concrete observer as we do it with time. We call it Relative Realities approach.

\section{Wigner's friend}\label{sectionWigner}

Wigner's friend approach to measurement is the first step towards the relative realities approach. When we are talking about measurement, we always need to refer to observer who is measuring. Similarly, strictly speaking ``Car is moving.'' is a nonsense, because we should rather say ``Car is moving relative to the ground''. We should always say ``Bob is measuring position of a particle'' instead of ``position of a particle is being measured''.

We already talked about measurements as entangling observer with the system from the observer's point of view. To illustrate Wigner's friend approach let us suppose we have a closed box with the dead-alive cat and Bob who does not like cats and is happy when he see the cat dead after he opens the box. He is in a normal mood before. So the whole system $cat + Bob$ is described by
\[
(\ket{\text{alive cat}}+\ket{\text{dead cat}})\ket{\neutranie}\ ,
\]
before opening the box and
\[\label{smileyfrownie}
\ket{\text{dead cat}}\ket{\smiley}+\ket{\text{alive cat}}\ket{\frownie}\ ,
\]
after opening the box.

From the point of view of Bob it is just a cat what he is observing. He cannot observe himself. But when some other observer came in and take look at the cat, let us call her Alice, she does not observe only the cat, but the whole Bell state (\ref{smileyfrownie}). He can look at Bob himself and see his sad face and because Alice know about Bob he hates cats she also knows the cat is alive. Or she can take look at alive cat and will know Bob is not smiling. The fact remains no matter what Alice does, she does not measure only the cat as Bob did, but the joint system $cat + Bob$. By measuring it, either by looking at Bob or at cat Alice entangle herself with this joint system. So the state (\ref{smileyfrownie}) is the system $cat + Bob$ from the point of view of outer observer, in our case Alice's view.

If we suppose Alice loves cats, for some another outer observer the system $Cat$+$Bob$+$Alice$ is then described by
\[
\ket{\text{dead cat}}\ket{\smiley}\ket{\frownie}_{A}+\ket{\text{alive cat}}\ket{\frownie}\ket{\smiley}_{A}\ .
\]

We can do the same for arbitrary amount of observers, each of them looking at something different. More people come to measure, more people entangle with the original system and observers who measured before.

Another consequence of this approach is quite strange and stunning discovery. When we have some measuring device and the system which is measured by it, the actual measurement from our point of view is not when the measuring device entangle itself with the system, but when we look at the device and entangle ourselves with it. If we have some device measuring dead-alive cat and we come take a look at the display whether it shows $0$ or $1$, the system we are observing is
\[
\ket{\text{dead cat}}\ket{0}+\ket{\text{alive cat}}\ket{1}\ .
\]
It does not matter how big the measuring device is, it is in the superposition anyway until we look at it, or until the information about dead-alive cat leaks from some other part of the device already entangled with the cat, for instance by emission of light.

How exactly is this process of entangling made is called measurement problem and is not solved yet entirely, however we try to put some insight into it in the last section of this chapter \ref{staringzenoeffect}.

\section{Brainwash experiment}

We talked already about presumption that macroscopic system behave in the very same way as the tiny one, since it is made of very same atoms and thus must obey the same laws of physics. Now we will use that assumption to construct brainwash experiment. It is nothing else than quantum eraser experiment with person acting the main role -- the role of observer whose knowledge is going to be erased. Let Alice be in that unpleasant position.

We suppose Alice has only one qubit spanned by $\{\ket{0},\ket{1}\}$ of memory. It is not much but will suffice to our cause. Now we will use her as an observer. For simplicity let us suppose she observes system (let us call it particle for now) entirely described by another qubit $\{\ket{L},\ket{R}\}$ (left, right). She observes this system or in other words she measures it but we already know that for some observer looking from the outside on the joint system of Alice and the particle it just behave like a unitary entangling operation between these two.

Suppose the initial state is
\[
\ket{\psi_i}=\frac{1}{\sqrt{2}}\ket{0}(\ket{L}+\ket{R})
\]
resulting in the state
\[
\ket{\psi_f}=\frac{1}{\sqrt{2}}(\ket{0}\ket{L}+\ket{1}\ket{R})=U\ket{\psi_i}
\]
after observation. One from the unitary operators able to do this is for example
\begin{equation}
\left( \begin{array}{cccc}
1 & 0 & 0 & 0 \\
0 & 0 & 0 & 1 \\
0 & 0 & 1 & 0 \\
0 & 1 & 0 & 0 \end{array} \right).
\end{equation}
in basis $\{\ket{0}\ket{L},\ket{0}\ket{R},\ket{1}\ket{L},\ket{1}\ket{R}\}$ but this operator is not uniquely defined. Some other unitary able to do this is for example
\begin{equation}\label{otherunitarychoice}
\left( \begin{array}{cccc}
0 & 1 & 0 & 0 \\
0 & 0 & \frac{1}{\sqrt{2}} & \frac{1}{\sqrt{2}} \\
0 & 0 & \frac{1}{\sqrt{2}} & -\frac{1}{\sqrt{2}} \\
1 & 0 & 0 & 0 \end{array} \right).
\end{equation}

The  evolution may describe Alice looking on one slit in double slit experiments or controlling one-path in our which-path experiment in section \ref{doublesliteraserSection} or opening the box with dead-alive cat or looking at a car whether it is parked on left or right side of a road.

Before an observation Alice know nothing about the system, but after it she knows whether the car is on the left $\ket{L}$ or on the right $\ket{R}$ and remembers it. If the car is on the left her brain remains in state $\ket{0}$ and when it is on the right her brain goes to state $\ket{1}$. For her it appears the car is either on the left side or on the right side, but for an outer observer both it true. We know both possibilities are ``real'' and we have to take them into account because otherwise particle in double-slit experiment would not interfere and two photons in Herzog et Al. Quantum eraser in section \ref{Herzogsection} neither.
The funny thing is we can eraser her memory and undo her observation, so she may be able to observe the car again. The only thing we need to do is find some unitary $\tilde{U}$ which applied to the final state gives us the initial one.
\[
\tilde{U}U\ket{\psi_i}=\ket{\psi_i}
\]
We know some certainly exist, for example we may put $\tilde{U}=U^{\dagger}=U^{-1}$ which actually acts as propagator taking back the time undoing observation to the time when it has never even happened. But this is not the only way how to do it. For example inverse of the unitary in (\ref{otherunitarychoice}) does the same thing and it does not even need to be propagating back in time. Another example of erasing device which does not need to propagate anything back in time is a Beam-splitter. Beam splitter acts as a simple Hadamard gate
\begin{equation}
\frac{1}{\sqrt{2}}\left( \begin{array}{cccc}
1 & 1 \\
1 & -1
\end{array} \right).
\end{equation}
and is its own inverse matrix so upon double passage everything goes back to the initial state.

The second unitary $\tilde{U}$ erase Alice's memory and now, when everything is back to the initial state, she can again look at the car and see it on the opposite side now. The unitary $\tilde{U}$ brain-washed her and that is why we call this brain-wash experiment. So she may have see the car on the left side for the first time and then after we brainwashed her she may see it on the right. Now we are getting to the true relative realities. For her in some time the car was on the left. It was her personal reality. But after we erased her memory, we also changed the possible reality she is able to see. In latter time she may see the car on the right side so her personal reality will be exactly the opposite. It seems we can change the realities as long as we erase memory of everyone who is involved and that is how any possible paradox is solved. No-one can say that in some earlier time his reality has been different, because to change the reality we need to erase his memory first.

We already constructed brain-washing machine but we can also do it differently, with more complexity. We can construct device who first take Alice's memories away and and erase it at a later time. For now suppose Alice has qutrit memory $\{\ket{0},\ket{1},\ket{2}\}$ and her initial state is $\ket{0}$ so she remembers whether she observed car already or not yet. Suppose we have another ``switching unit'' system consisted of $\{\ket{u},\ket{d}\}$ (up, down). Now we can temporarily move Alice's memories to switching unit, switching unit moves to the car, disentangles and everything is back to the initial state.
\[
\begin{split}
&\frac{1}{2}\ket{0}(\ket{L}+\ket{R})(\ket{u}+\ket{d})\\
\stackrel{U_1}{\longrightarrow} &\frac{1}{2}(\ket{1}\ket{L}+\ket{2}\ket{R})(\ket{u}+\ket{d})\\
\stackrel{U_2}{\longrightarrow} &\frac{1}{\sqrt{2}}\ket{0}(\ket{L}\ket{u}+\ket{R}\ket{d})\\
\stackrel{U_3}{\longrightarrow} &\frac{1}{2}\ket{0}(\ket{L}+\ket{R})(\ket{u}+\ket{d})
\end{split}
\]

One can easily find some unitary operators achieving it, an example is in appendix \ref{unitary3}. The same switching device could be used to move Alice's memory to Bob. Note we cannot just erase the quantum information (no-deleting theorem \cite{PatiNoDeleting}) nor to copy it in general (no-cloning theorem \cite{NielsenChuang}).

Nevertheless, although this may seem revolutionary, it may be very hard to accomplish such experiment. It is mainly due to the continuous interacting with other systems and decoherence of involved car and we already know to perform the successful eraser experiment we need to involve every particle interacting with a car. This may be overcome by using very small systems instead of a whole car, for example a single particle, but still there is a problem how to accomplish the proper erasing unitary $\tilde{U}$. The interaction has to be made through some medium otherwise Alice would not see anything. For example by photon de-excitation. But then the erasing unitary would represent pulling the photon back from Alice's eye and stuffing it back the the observed particle. Also no-one could tell anyone whether the experiment about relative realities has been successful because to make it successful memories of all participants have to be erased. Usefulness is also questionable, nevertheless, when you have a chance to save the life of a cat, why not to utilize it?

\section{Relative Realities}\label{sectionRelative Realities}

This chapter will be motivated by a quantum eraser experiment in section \ref{Herzogsection}. Let us reintroduce the this experiment depicted on figure \ref{Herzog1}. In this experiment we had high energy photon which transformed into two low energy photons. Twice. At the first passage there was some probability the consecutive pair of lower energy photons are created and some probability that they are not created and high energy photon passes through. After it is reflected back and has the second chance to create the same pair of photons. However, at the end we see the interference between pair created upon first passage and the pair created upon second. So both of these possibilities of creation were actually true. There also must have been time where both high-energy photon and the created pair coexisted, otherwise the system would not interfere. And it interferes. This is not just thought experiment but a real one with well documented interference pattern \cite{ComplementarityHerzog}. So this experiment seems to violate conversation energy law, since at the beginning we have one high-energy photon and some later time we have this photon + another pair yielding the same energy. In fact, the law of energy conversation has quite different meaning here. If we take a look at arbitrary part of the system, we never see both high-energy photon and a pair at the same time. Only one possibility comes true. Also, as we showed in section \ref{sectionComplementarityofinterferenceandcorrelations} this observing prevents system from interfering. By observing it we banned the second possibility and thus also saved the conservation energy law. At a given time we may never see any paradox because we see either high energy photon or the pair. We can never see both at once. Conversation energy law here just says the mean value of energy is conserved and when there is a three percent chance the pair is produced upon one passage, also this possibility possess only three percent of energy, while the other possibility yields the rest 97 percent.

In the previous section we also showed in principle there is a way how to take everything back. If Alice look at this experiment trying to determine whether she see only one high-energy photon or one-pair of low energy photons only one possibility comes true. But by applying some unitary operator on the whole system including Alice we may able to take everything back. Then she may look again. She may have seen high-energy photon for the first time, but for the second time she may see a pair of low energy photons.

So even the existence of particle is relative. From the point of view of some outer observer (Bob for example) the system behaves nicely by some unitary evolution and can show interference, but for the observer stuck inside the system itself (Alice) it appears very differently -- it seems only one reality (history, possibility) is true.

For Alice being in the system and observing it means the system does not evolve unitary anymore. It is because she interacts with the system and getting entangled with it, this is usually called measurement. In Everett's Many World interpretation we assume in every measurement the world splits into more worlds with certain probability where the consecutive worlds differ in the measurement outcome. Thus the whole evolution of the universe is consisted of infinite and irreversible splitting. Nevertheless in previous section we proved we can go back from one such world to another by a simple quantum erasure. Also measurement is strictly local issue, since it does involve only people interacting with the observed system, not the people far away who have nothing to do with it.

In the relative realities approach we suggest there is only one world. The world consisted of every possibility where these possibilities interfere with each other. However by looking at this world from the inside, only some possibilities come true. By looking at this world we build our own reality\footnote{And we will not discuss whether we have any control over how it is built or not, whether we can choose the outcome of our measurement or not.} while interacting with it. However this reality we build is not irreversible. Someone from outside can change for us by a quantum erasure and give us a second chance to observe.

\begin{figure}[!h]
\begin{center}
\includegraphics[width=0.9\textwidth]{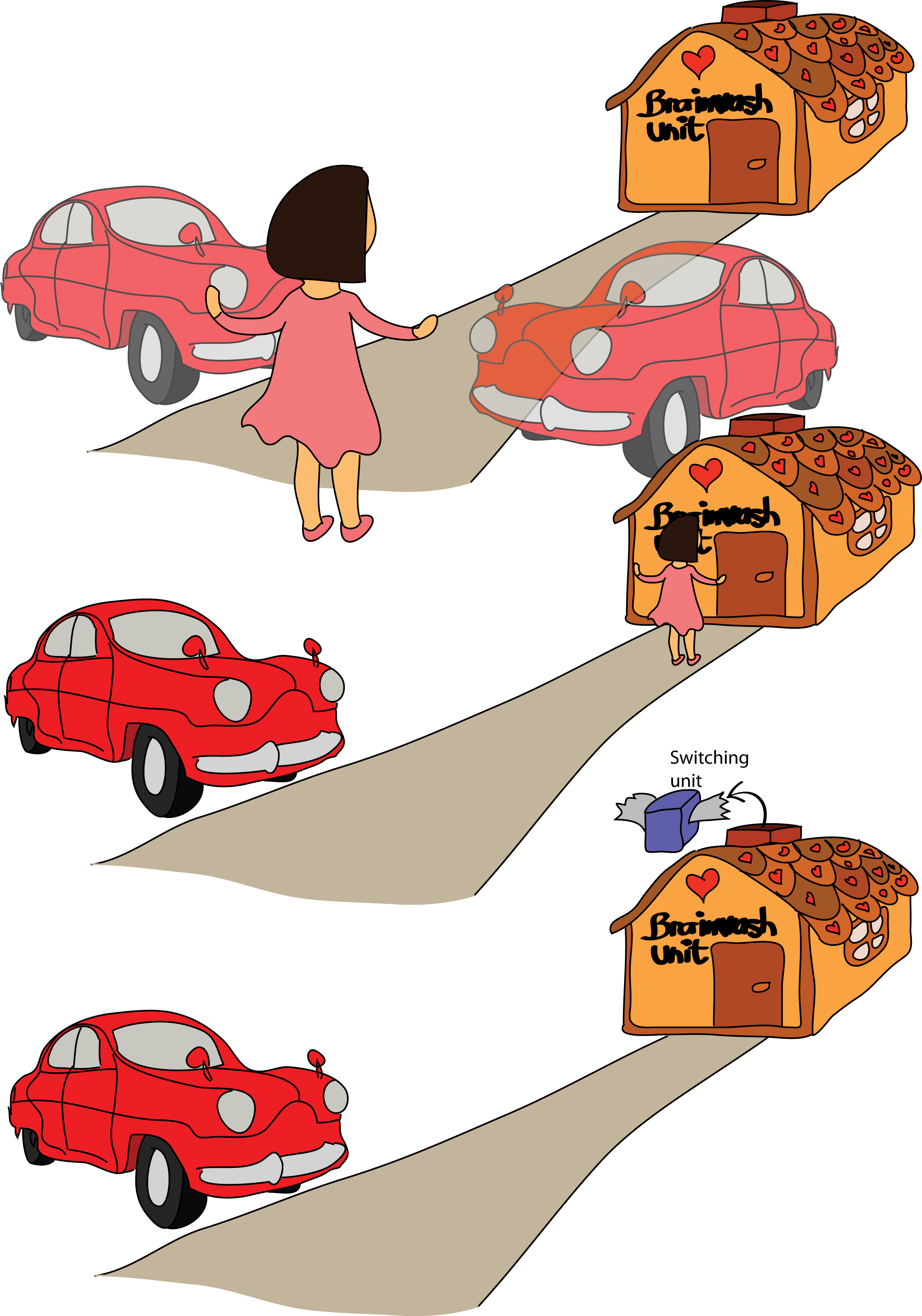}
\caption{Alice in Wonderland. Alice look at the car and see it on the left side. She does not like it because it is much easier for her to ride from the right side. Thus she decides to go to brainwash unit. Her knowledge about parked car has been passed to the switching unit.}\label{aliceinwonderland4}
\end{center}
\end{figure}

\begin{figure}[!h]
\begin{center}
\includegraphics[width=0.9\textwidth]{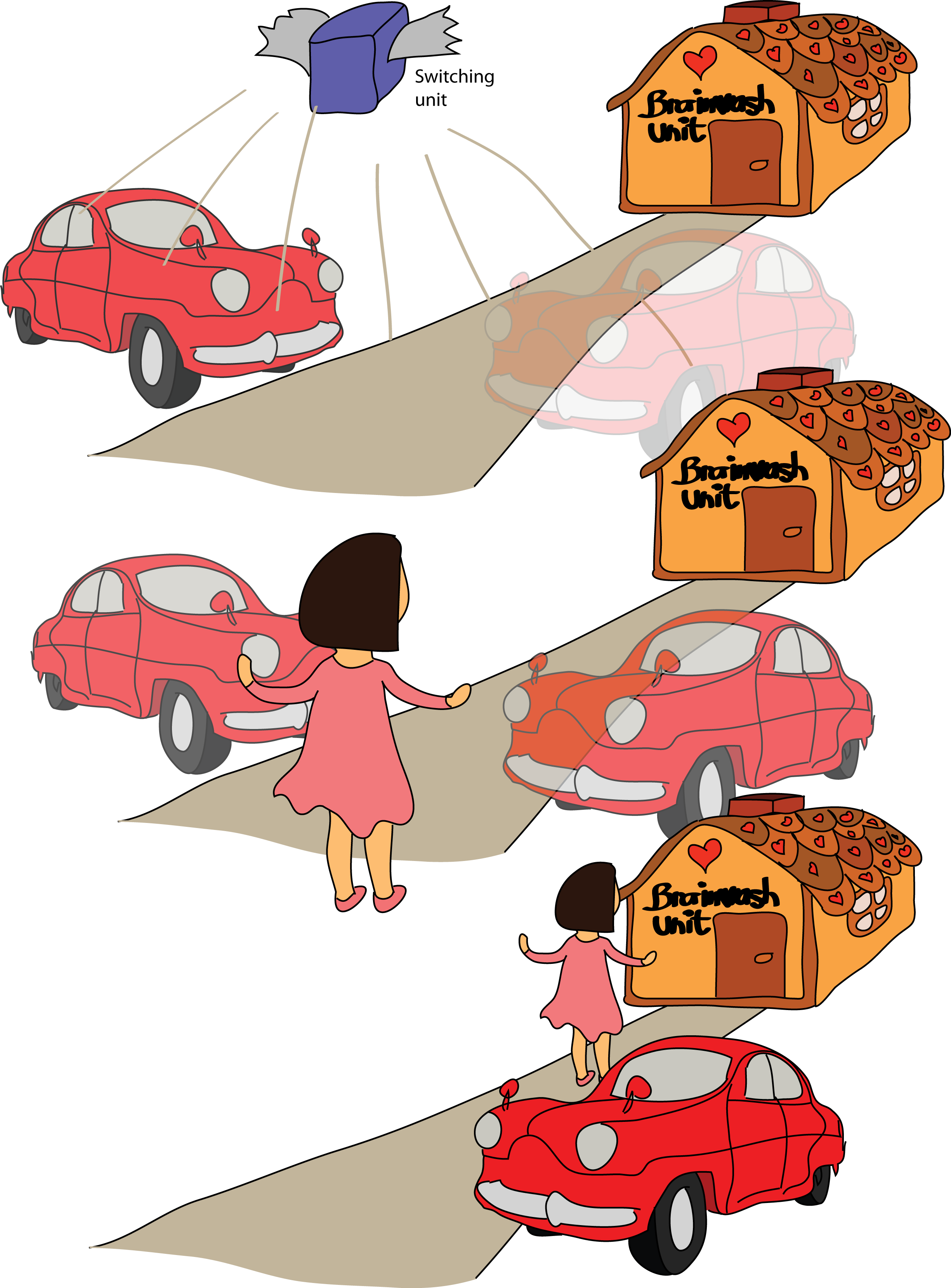}
\caption{Alice in Wonderland. Switching unit flies back to the parked car and makes everything undecided again. Alice can take a look for the second time. She sees the car on the right side now.}\label{aliceinwonderland5}
\end{center}
\end{figure}

\begin{figure}[!h]
\begin{center}
\includegraphics[width=0.8\textwidth]{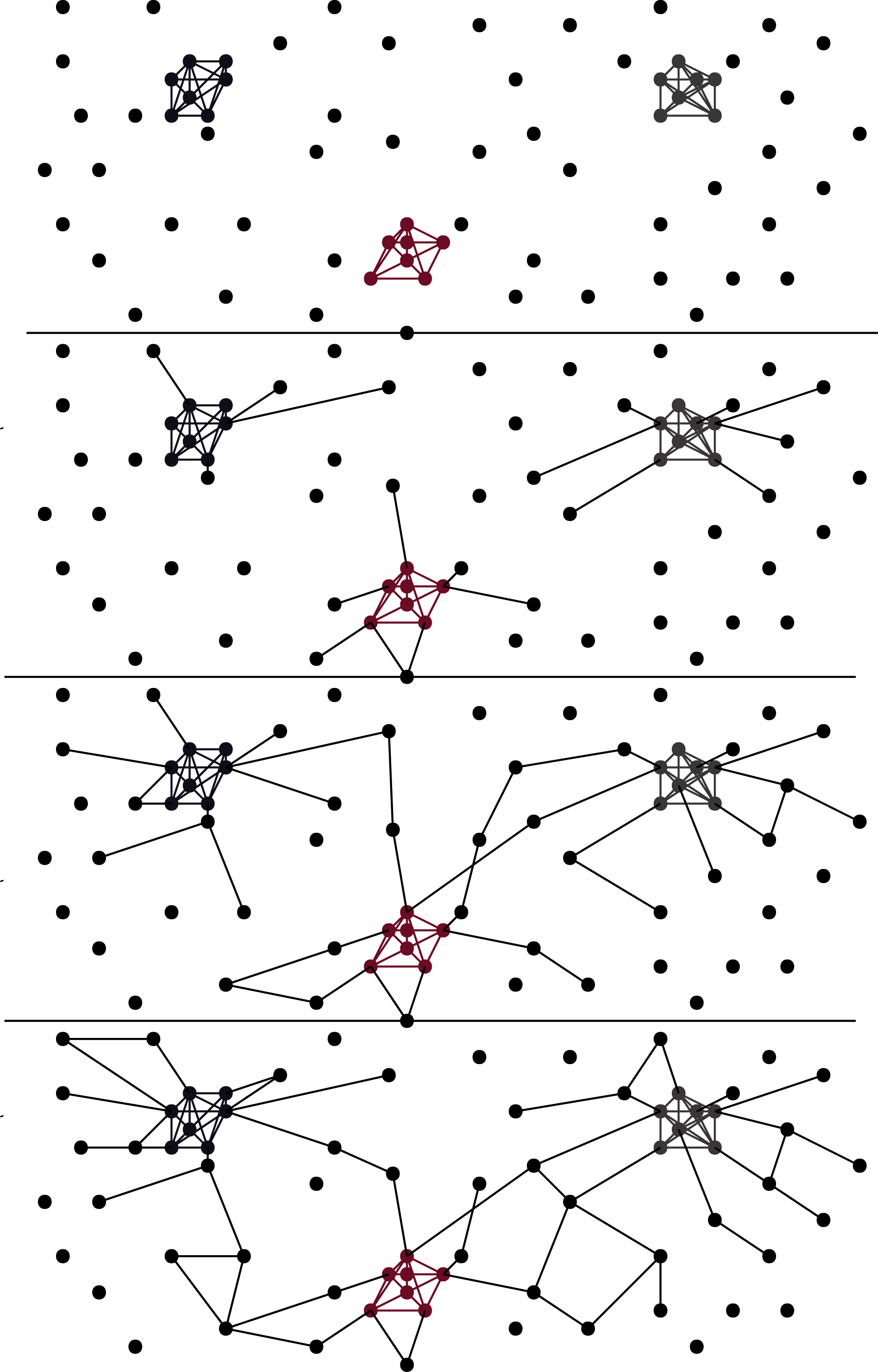}
\caption{Personal (relative) realities of three people (highly entangled and continuously interacting systems). As these people interact with world, observe it, they get entangled with it. As shown in Delayed Choice Entanglement Swapping, they can also entangle indirectly, they can be entangled with things they never met before. That is why we chose only black lines to represent entanglement. Amount and specific type of entanglement depends on a type of specific interaction. On the last picture we also depicted two quantum erasures.}\label{personalrealities}
\end{center}
\end{figure}

We illustrate it in the following comics ``Alice in Wonderland'' (figure \ref{aliceinwonderland4} and \ref{aliceinwonderland5}) and Personal realities spreading (figure \ref{personalrealities}) at the end of this chapter.

\section{Continuous evolution and Quantum Zeno effect}

Another problem in Many World interpretation and thus also Relative realities approach we intentionally avoided until now is the problem with continuously evolving system, when observer continuously entangles with its observed system. Until now we worked with unitary evolution independent of time, represented by some unitary jump (appendix \ref{listofUnitaries}) instead of rather unitary evolution continuously transformed from identity operator. But a lot of unitary evolution is time-dependent. Imagine for example Alice slowly opening the box with dead-alive cat or Bob looking at the decaying atom.

Suppose Bob is happy when sees the the alpha particle from the atom decay because after that he can go finally home. Bob with the atom is described then by
\[\label{continuousBob}
\sqrt{e^{-\lambda t}}\ket{U}\ket{\frownie}+\sqrt{1-e^{-\lambda t}}\ket{Th}\ket{\smiley}\ .
\]

If we suppose initial state
\[
\ket{U}\ket{\frownie}
\]
the unitary matrix between initial and final state is then for example
\begin{equation}
\left( \begin{array}{cccc}
\sqrt{e^{-\lambda t}} & \sqrt{1-e^{-\lambda t}} & 0 & 0 \\
0 & 0 & 0 & 1 \\
0 & 0 & 1 & 0 \\
\sqrt{1-e^{-\lambda t}} & -\sqrt{e^{-\lambda t}} & 0 & 0 \end{array} \right),
\end{equation}
in basis $\{\ket{U}\ket{\frownie},\ket{U}\ket{\smiley},\ket{Th}\ket{\frownie},\ket{Th}\ket{\smiley}\}$.

Alice continuously opening the box can be described for instance by continuous evolution
\[\label{continuousAlice}
\begin{split}
\frac{e^{-it}}{\sqrt{2}}&(\cos t \ket{\text{alive}}\ket{\neutranie}+\cos t \ket{\text{dead}}\ket{\neutranie}\\
+&i\sin t \ket{\text{dead}}\ket{\smiley}+i\sin t \ket{\text{alive}}\ket{\frownie})
\end{split}
\]
with initial state
\[
(\ket{\text{alive}}+\ket{\text{dead}})\ket{\neutranie}\ ,
\]
at $t=0$ and final state
\[\label{smileyfrownie}
\ket{\text{dead cat}}\ket{\smiley}+\ket{\text{alive cat}}\ket{\frownie}\ ,
\]
at $t=\frac{\pi}{2}$, which is the very same as presented in section \ref{sectionWigner}. The unitary matrix able to do that is written in appendix \ref{unitary4}.

Both of these examples shows the usual interpretation of splitting into two worlds with some probabilities seems somehow irrelevant. How could Alice or Bob enter different worlds with certain probabilities when these probabilities change continuously? Or do they enter just once and then stay in this new world? When exactly do they jump into their new world? Or do they keep jumping from one world to another?

All these question arise because we tried to describe what Alice and Bob see using the knowledge of some another outer observer. But states (\ref{continuousBob}),(\ref{continuousAlice}) do not describe what Bob and Alice sees, it does describe what the outer observer observing both Bob and atom or Alice and Cat will see when he decides to look at them. The question is, is there any relationship about these states and about what Bob and Alice actually see? Can we derive when Bob will detect an alpha particle and when Alice finally see the dead or alive cat just from this outer observer's wave function?

One example which shows there is a difference about what outer observer and inner observer see is Quantum Zeno effect. Simply speaking it says very frequent measurements\footnote{This is usually referred to as {\em continuous measurement}, but it is always large but finite amount of measurements with small time lags between measurements.} of the same system inhibits evolution of this system. This is mainly due to the fact that wave function quantum systems change as $\sim \Delta t$ while probability of measuring the state different from the one measured in the previous measurement is proportional to $\sim (\Delta t)^2$, where $\Delta t$ is a delay between two consecutive measurements. Probability the state remains the same between two consecutive measurements is then
\[
P=1-\frac{(\Delta t)^2}{\Delta t}
\]
which goes to $1$ when the delay is sufficiently small.

This effect has been experimentally proved, for example in \cite{ItanoZenoEffectExperimental} where authors of the article observed this effect in an rf transition between two 9Be+ ground-state hyperfine levels. Short pulses of light, applied at the same time as the rf field, made the measurements. There are also theoretical works, for example on a particle in double-well potential \cite{GagenContinuousPositionMeasurements}, where authors prove frequent position measurement of particle in such potential inhibits tunneling from one well to the second. However, all of these quantum Zeno effects require \emph{active} measurements, where experimenter keep pumping some sort of energy into the system. It is the laser beam in the first example which keeps electrons jumping from low orbit to higher one and back by spontaneous de-excitation and in the second example the position measurements keep pushing the particle to the higher and higher energy level which finally leads to higher energy than the barrier between the two wells and sets the particle free.

However, such Quantum Zeno effect does not solve our problem. Because in our problem we do not put any energy in, we just passively look around. It is a simple fact just looking at some unstable material we are not able to inhibit its decay.\footnote{I experienced it myself. Holding Americium 243 in my hand and looking at it has not prevented my Geiger--M\"{u}ller counter from clicking.} To solve our problem, to answer what observers are experiencing we need something else. We introduce it in the next section.

\section{Possible solution to what Observers are experiencing, the true form of wave-particle duality, passive Zeno effect}\label{staringzenoeffect}

Wave-particle duality is usually explained as complementarity of wave and particle behavior, in other words, that both wave-like and particle-like behavior cannot be observed simultaneously. As some experiments show \cite{KaiserDelayedChoiceExperimental} and as we showed in the first chapter \ref{doublesliteraserSection} there is a continuous transformation between wave-like and particle-like behavior.

One very simple example is double-slit experiment. Normally when we send some photons through double slit it interferes and we will see interference fringes on our interference plate -- we have wave-like behavior. But when we put correctly oriented half-wave plate in front of one slit, it turns the polarization of light coming through this slit perpendicular to the polarization of light coming through the second slit and the interference disappear -- we have particle-like behavior. Well but we can do also something between. We can orient half-wave plate on the first slit in a such way it turns the polarization just by $45^\circ$. Then we will see less visible interference pattern with some scent of non-interference intensity peaks in front of each slit -- we have something between particle-like and wave-like behavior. We could do the same calculations we did in the second chapter \ref{sectionComplementarityofinterferenceandcorrelations} when deriving complementarity of correlation and one particle interference. But all the calculation were made using waves, not particles. Note that particle behavior comes out only when we measure. As in Tonomura's double-slit experiment using electrons, the interference pattern was not seen at once, but has to be filled by single hits finally forming the well-known fringes. Thus we suggest the true form of wave-particle duality:

\bigskip
\textbf{Particles always evolve as waves, their particle nature comes out only when we entangle with them, measuring them. When we describe the system we do not interact with, particles come through every possible history and each of these histories are real, each of them contributes to some measurable quantities. Quantization is a result of entangling ourselves with the system. When we interact with the system or, in other words, we measure the system, we see only quanta - something we call particles. Wave behavior thus refers to the system we do not interact with, while particle behavior refers to our observation of the system.}
\bigskip

Now we can finally answer what Bob and Alice see. While from the outside is Bob described by
\[\label{continuousBob2}
\sqrt{e^{-\lambda t}}\ket{U}\ket{\frownie}+\sqrt{1-e^{-\lambda t}}\ket{Th}\ket{\smiley}\ .
\]
His knowledge about the state change must be carried by some interaction particle, some quanta he is able to detect. In this case it is an alpha particle, which gives the information about the atom decay. From the view of outer observer the system Bob+atom is evolving continuously (wave behavior), but from the point of view of Bob it is strictly jump effect. He just detects an alpha particle at some point. To derive when he detects the particle we use the wave function describing the overall state (\ref{continuousBob2}). From the form of the wave function we can be sure at time $t_0=0$ the atom has not decayed yet. The probability it decays between time $(0,t)$ is equal to \footnote{U as uranium, Th as thorium.}
\[
p(U\rightarrow Th(0,t)|U(t_0=0))=\left(\sqrt{1-e^{-\lambda t}}\right)^2=1-e^{-\lambda t},
\]
which is the very same probability Bob detects an alpha particle between $(0,t)$. Let us denote probability he detects particle at time $\tau$ as
\[
p(U\rightarrow Th(\tau)|U(t_0=0))=\rho(U\rightarrow Th(\tau)|U(t_0=0))\mathrm{d}\tau\ ,
\]
where $\rho(\cdot)$ is probability density. From the construction it is obvious that
\[
\int_0^t\rho(U\rightarrow Th(\tau)|U(t_0=0))\mathrm{d}\tau=p(U\rightarrow Th(0,t)|U(t_0=0))\ .
\]

After differentiation we get the differential equation for $\rho$
\[
\rho(U\rightarrow Th(t)|U(t_0=0))=\frac{\mathrm{d}p(U\rightarrow Th(0,t)|U(t_0=0))}{\mathrm{d}t}=\lambda e^{-\lambda t}.
\]

So we derived Bob will detect incoming particle at time $t$ with probability
\[
p(U\rightarrow Th(t)|U(t_0=0))=\lambda e^{-\lambda t}\mathrm{d}t\ .
\]

From general wave function describing both Bob and atom we were able to derive what and when Bob is going to see. Now back to the relativity of realities. For outer observer looking\footnote{Strictly speaking looking is a wrong word here. Outer observer should not look, because looking usually refers to some sort of measurement, some type of the interaction with the system. To be precise, we should rather talk about observer not looking but describing the system Bob+atom.} at both Bob and atom the whole system is in superposition, while for Bob the atom stays in the same state $\ket{U}$ until he detects an alpha particle. After detection it stays in $\ket{Th}$. Now we illustrated the abysmal difference between observer describing the system while not interacting with it and observer interacting with the system. While for the outer observer the evolution is continuous transformation, for inner observer the evolution is a jump process.\footnote{As reader probably noticed we just paved our path to well-known Born rule.}

Now suppose Bob has not detected an alpha particle up until $t_0\>0$. Knowing this we can modify his probability density to better correspond the current knowledge. Now the differential equation is
\[
\rho(U\rightarrow Th(t)|U(t_0))=\frac{\mathrm{d}p(U\rightarrow Th(0,t)|U(t_0))}{\mathrm{d}t}=\lambda e^{-\lambda (t-t_0)}
\]
This is quite interesting. Not observing anything gives Bob some additional information about the system and he is able to modify his probabilities accordingly. Not only detecting a particle but also not detecting anything gives Bob some information and his wave function describing his observed system changes to the conditional one.

Now we can also compute probability of detecting a particle in a very close future, while not detected before. For $\Delta t=t-t_0\ll 1$
\[
p(U\rightarrow Th(t)|U(t_0))=\lambda e^{-\lambda \Delta t}\Delta t\doteq(1-\lambda \Delta t)\lambda \Delta t\doteq\lambda \Delta t\ .
\]

Now we get back to Alice's observation of alive-dead cat. Namely wave function
\[\label{continuousAlice2}
\begin{split}
\frac{e^{-it}}{\sqrt{2}}&(\cos t \ket{\text{alive}}\ket{\neutranie}+\cos t \ket{\text{dead}}\ket{\neutranie}\\
+&i\sin t \ket{\text{dead}}\ket{\smiley}+i\sin t \ket{\text{alive}}\ket{\frownie})\ .
\end{split}
\]
In Bob's case we had some particle which tells us the state has changed. We need to have some here too. For that reason suppose the dead cat turns blue and thus emits blue light, while alive cat remains ginger and emits red light. Now when Alice opens the box seeing color of the first photon she can tell whether the cat is alive and when is dead. After receiving this first photon Alice will be receiving many others of the same color, because the original Quantum Zeno effect takes place. Simply speaking, if cat proves to be alive, there is much greater chance it will stay alive if the photon emission is frequent enough. If cat proves to be dead, it will be most likely dead some time later.

Now state $\ket{\text{alive}}\ket{\neutranie}$ refers to the situation where the cat is alive but we did not received our photon yet so we do not know yet. Similarly with the state $\ket{\text{dead}}\ket{\neutranie}$. Finally the states $\ket{\text{alive}}\ket{\frownie})$ and $\ket{\text{dead}}\ket{\smiley}$ refer to the cases when we received the photon and found out whether the cat is alive or dead.

Analogously to the Bob's case, probability of finding out in interval $(0,t)$ the cat is dead with initial state
\[
\ket{i}=\frac{1}{\sqrt{2}}(\ket{\text{alive}}+\ket{\text{dead}})
\]
when slowly opening a box is
\[
p(blue(0,t)|\ket{i}(t_0=0))=\frac{e^{-it}}{\sqrt{2}}(i \sin t)\frac{e^{it}}{\sqrt{2}}(-i \sin t)=\frac{1}{2}\sin^2 t
\]
and the respective probability density of finding out the cat is dead at time $t$ is
\[
\begin{split}
\rho(blue(t)|\ket{i}(t_0=0))&=\frac{\mathrm{d}p(blue(0,t)|\ket{i}(t_0=0))}{\mathrm{d}t}\\
&=\sin t \cos t=\frac{1}{2}\sin (2t)\ .
\end{split}
\]
We get the same probability density when calculating alive cat
\[
\rho(red(t)|\ket{i}(t_0=0))=\frac{1}{2}\sin (2t)\ .
\]
So at any time $0\leq t\leq\frac{\pi}{2}$ the probability we find out the cat dead or alive is the same.

We know that if some outer observer look at Alice after $t>\frac{\pi}{2}$ he must find out Alice either happy or sad, nothing between. That simply means probability of detecting blue photon in the next short interval $(t_0,t)$, $\Delta t=t-t_0$ supposing we have not detected any particle until $t_0$ must go to infinity as $t_0$ goes to $\frac{\pi}{2}$. Also not detecting a particle until $t_0$ means that we just have to throw out some possible outcomes from our sample space and renormalize the probability of the others. We throw out all outcomes of type ``Alice detected a particle at time $\tau<t_0$''. Probability density of detecting blue photon at time $t>t_0$ while Alice has not detected any particle until $t_0$ is thus\footnote{We could have use the same reasoning for the Bob and his unstable atom and get the same result.}
\[
\begin{split}
\rho(blue(t)|\ket{i}(t_0))&=\frac{\rho(blue(t)|\ket{i}(0))}{\int_{t_0}^\frac{\pi}{2}\rho(blue(\tau)|\ket{i}(0))\mathrm{d}\tau+\int_{t_0}^\frac{\pi}{2}\rho(red(\tau)|\ket{i}(0))\mathrm{d}\tau}\\
&=\frac{\frac{1}{2}\sin (2t)}{2\int_{t_0}^\frac{\pi}{2}\frac{1}{2}\sin (2\tau)\mathrm{d}\tau} =\frac{\sin(2t)}{1+\cos(2t_0)}\ .
\end{split}
\]

After short derivation we get the probability of detecting blue photon in the next short interval $(t_0,t)$, $t_0<t<\frac{\pi}{2}$ supposing we have not detected anything until $t_0$\footnote{The reader may noticed that the probability has not correct dimension. It is because we work only win $\sin(t)$ instead of $\sin(\omega t)$.}
\[
p(blue(t)|\ket{i}(t_0))=\frac{\sin(2t)}{1+\cos(2t_0)}\Delta t\doteq\frac{\sin(2t_0)+\cos(2t_0)\Delta t}{1+\cos(2t_0)}\Delta t\ .
\]

For $t_0\approx\frac{\pi}{2}$ using Taylor series we get
\[
p(blue(t)|\ket{i}(t_0))=\frac{\Delta t^2}{2(\frac{\pi}{2}-t_0)^2}\ .
\]

Note that the actual state of the cat does not matter on Alice's view anymore. All she can tell about the cat is mediated by the particle she is receiving. She supposes the cat is dead when she received the blue photon and that is nearly all she can do. Whether the blue photon really tells her that is totaly different question. It just seems we are bounded by our limited ability to perceive and all we suggest about surrounding world is just an interpretation of little information we receive. What we see is not anymore about the actual object but rather about particles mediating the interaction. Blue photon just makes Alice happy and red photon makes her sad. However these mediating particles yield some information about the object they have been emitted from and that is why we can say something about it.

\begin{figure}[!h]
\begin{center}
\includegraphics[width=1\textwidth]{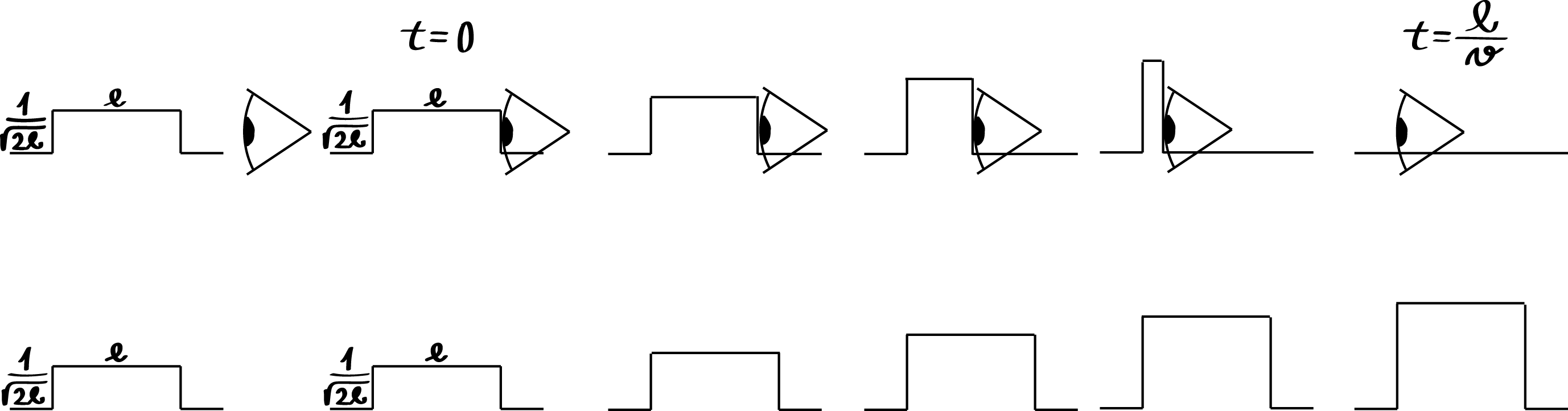}
\caption{As we look on one part of the system and we do not receive anything probability of particle being somewhere else is rising. If we do not detect anything we can be sure particle is somewhere on the other side.}\label{passiveZeno}
\end{center}
\end{figure}

To illustrate better passive Zeno effect let us present our last example in this thesis. Imagine particle going through 2 paths at once, while Bob controls only one path. It is just like looking into one of two slits, with one slight difference, we presume quite idealized rectangular wave packet now. Suppose even though Bob looks into one slit he will not detect any particle, the particle just went through the other way. Illustration is on figure \ref{passiveZeno}.

If we omit from our reasoning the actual horizonal position of a particle and rather focus on what Bob is seeing, the wave function is
\[
\sqrt{\frac{1}{2}\left(1-\frac{t}{T}\right)}\ket{u}\ket{0}+\sqrt{\frac{1}{2}\frac{t}{T}}\ket{u}\ket{1}+\sqrt{\frac{1}{2}}\ket{d}\ket{0}\ ,
\]
where $T=\frac{l}{v}$ is duration of interaction, $l$ initial length of wave packet, $v$ wave packet velocity, $\ket{u},\ket{d}$ particle position which refers to up or down, $\ket{0},\ket{1}$ excitation of Bob's eye, $0$ refers to non-excited eye (not particle detected) and $1$ to excited eye (particle detected).

The interesting fact about this experiment is even though Bob did not detect anything, he entangled himself with the system and destroyed possible interference. He did not notice any change, still there was a physical consequence. That is what we call passive Zeno effect. Because similarly to the original quantum Zeno effect the wave function collapsed, but now it was not a consequence of active contribution of an observer, it was rather amusing circumstance, Bob did not even notice he achieved something like that.

Since this experiment is mathematically somewhat different from the previous examples, let us calculate the respective probabilities for the last time.

With initial state
\[
(\ket{u}+\ket{d})\ket{0}
\]
the probability density of detecting a particle at time $t$, $0<t<T$
\[
\begin{split}
\rho(0\rightarrow 1(t)|\ket{i}(t_0=0))&=\frac{\mathrm{d}p(0\rightarrow 1(0,t)|\ket{i}(t_0=0))}{\mathrm{d}t}\\
&=\frac{1}{2T}\ .
\end{split}
\]
Probability density of detecting particle at time $t>t_0$ while Bob has not detected anything until $t_0$ is
\[
\begin{split}
\rho(0\rightarrow 1(t)|\ket{i}(t_0))&=\frac{\rho(0\rightarrow 1(t)|\ket{i}(0))}{1-p(0\leq t\leq t_0|\ket{i}(0))}\\
&=\frac{\frac{1}{2T}}{1-\int_{0}^{t_0}\frac{1}{2T}\mathrm{d}\tau}=\frac{1}{2T-t_0}\ ,
\end{split}
\]
where the denominator $1-p(0<t<t_0|\ket{i}(0))$ is equal to sum of probabilities of possible outcomes on the presumption Bob has not detected anything until $t_0$. As before, we just renormalized probabilities of possible events in our new sample space.
Probability of finding a particle in the next short interval $(t_0,t)$ when Bob has not detected any particle until $t_0$ is then
\[
P(0\rightarrow 1(t)|\ket{i}(t_0))=\frac{1}{2T-t_0}\Delta t\ .
\]

To show renormalization of probabilities is a right way of calculating conditional probabilities, let us do it on the atom decay again, where $\rho(U\rightarrow Th(t)|U(0))=\lambda e^{-\lambda t}$. Our new conditional probability of detecting alpha particle on the presumption Bob has not detected it until $t_0$ is then
\[
\begin{split}
\rho(U\rightarrow Th(t)|U(t_0))&=\frac{\lambda e^{-\lambda t}}{\int_{t_0}^\infty\lambda e^{-\lambda \tau}\mathrm{d}\tau}=
\frac{\lambda e^{-\lambda t}}{1-\int_{0}^{t_0}\lambda e^{-\lambda \tau}\mathrm{d}\tau}\\
&=\frac{\lambda e^{-\lambda t}}{e^{-\lambda t_0}}=\lambda e^{-\lambda (t-t_0)}\ ,
\end{split}
\]
which is the very same result we received before directly from differentiation of $p(U\rightarrow Th(0,t)|U(t_0))$.

At the end of this chapter let us summarize what we learned here.

\bigskip
\textbf{For outer observer who do not interact with the system system transforms continuously through some unitary evolution, without any jumps, while observer interacting with the system never see any continuous transformation. Instead the information he receives about the system is delivered to him by quanta -- indivisible amounts of energy. Probabilities of receiving this quanta at a time $t$ can be calculated by differentiating squared coefficients of the wave function describing both observer and the system he interacts with.}
\bigskip

We added the presumption that for outer observer there is always continuous transformation, without any jumps. You can immediately argue that for example applying unitary matrix
\begin{equation}\label{jumpunitary}
U=\left( \begin{array}{cc}
0 & 1  \\
1 & 0  \end{array} \right).
\end{equation}
is not a continuous transformation. But using Occam's razor we suggest all of such jumping unitary matrices are limits of continuous ones, for example (\ref{jumpunitary}) being limit of
\begin{equation}
U=e^{-i\omega t}\left( \begin{array}{cc}
\cos (\omega t) & i\sin(\omega t)  \\
i\sin(\omega t) & \cos (\omega t)  \end{array} \right).
\end{equation}
for $\omega\rightarrow\infty$, so the continuous transformation is so fast it seems as an instant.
That is why we said ``For outer observer the evolution is always continuous.'' ``For inner (interacting) observer the evolution is always jumping.''

\chapter{Conclusion}

We began this thesis by presumption ``Observer is just another quantum system'' and later added ``Close system evolves through unitary interaction''. These two presumptions accompanied us on our way to understanding.

Using it, we were able to mathematically describe why obtainable which-path information prevents system from interfering, we found that there is a continuous transformation from wave-like behavior and particle-like behavior.

Using these two presumption we were able to prove why Free will test, the experiment which tests one's free will based on the quantum entanglement is not possible. We were able to prove we cannot predict the future using quantum entanglement alone. That was the Non-communication theorem.

Using these two presumptions we were able to prove formerly experimental concept of complementarity of one-particle interference and the correlations. We proved that better correlation leads to worse visible interference pattern and vice versa and we found a system which exhibits both interference and correlations to some other system. This concept was mathematically the very same as the proof of why which-path information prevents system from interfering. It is basically the same phenomenon.

These two presumptions are the basic building blocks of the Many World interpretation. With the knowledge gained during studies of quantum eraser experiments we used this interpretation as a start point to more complex view, Relative Realities approach. We proved there is a way how to pass between two worlds, which is the opposite of what original Many World interpretation suggests. Still to do it, we need to erase every knowledge about the former world first to prevent paradoxes. We showed there is a different reality for each observer. At different time people can see the same thing differently. If we erase someone's memory, he may be able to do the same measurement for the second time and get a different result. Still, no paradox occurs since the involved person does not remember anything from his former experience. Also two people cannot see a different result on the same experiment, because the latter do not measure only the system, but both system and the person who measured first.

In Relative realities approach, we suggest there is only one world, full of all possible superpositions and all possible outcomes. By observing it we entangle ourselves with it and reduce our scope. Nevertheless, there is coming back by disentangling ourselves using quantum erasure. Thus in contrast to Many World interpretation, this approach fits better the time symmetric nature of Schrödinger's equation.

At last, we showed probability calculation in Many World interpretation does not fit very well with time-dependent interaction between the system and the observer. We tried to fix this problem and we suggested a solution. While for the ``outer'' observer who do not interact with the joint system of ``inner'' observer and system ``inner'' observer interacts with the evolution is always continuous, for ``inner'' observer interacting with the system the evolution is always jump process. The information about observed system is delivered to him by quanta -- indivisible package of energy (or information). Probability of receiving quanta at a given time $t$ can be calculated from universal wave function\footnote{Here we mean the wave function that describes both the ``inner'' observer and the ``observed system'' itself  from the point of view of the ``outer'' (or reference) observer.} by differentiating squared coefficients belonging to observer's state. Calculating conditional probabilities of type ``Bob will detect a particle at a time $t$ while he did not detected anything until now'' is done by renormalizing of the remaining probabilities, or in other words, by renormalizing probabilities which refer to remaining events from our sample space. We successfully used this computation to describe Bob observing an atom decay and calculated two other interactions, namely Alice slowly opening a box with the dead-alive cat and Bob looking on one of two slits. We also introduced passive Zeno effect, which says an observer can destroy possible interference pattern even though he did not notice any change, he did not registered anything. That was our small contribution to the measurement problem.

\appendix

\chapter{Time evolution of some special systems}\label{timeEvolution}

Here we will show that the systems where each particle has only one energy level and these particles do not longer interact time ordering of measurements on exclusive parts (different particles) has no particular importance and thus we can omit it in our derivations. All of the experiments in the first chapter satisfy such presumption. For example in Wheeler's quantum eraser it does not matter whether the photon goes through the upper or lower part, since it energy is determined only by its wavelength and it is the same in both cases. In Free will experiment the reasoning is the same. In Delayed Choice Entanglement Swapping we do not measure position but polarization instead. In this special case it does not matter, but if we took electrons with spins and we did the same experiment in magnetic field, it would have behave differently, since the energy of spin up would be different than energy of spin down.

Suppose we have $n$ non-interacting particles and each can have only one value of energy. The governing hamiltonian is then just a multiple of identical operator
\[
H=H_1+H_2+\cdots+H_n
\]
where $$H_1=E_1\bold{1}\otimes\bold{1}\otimes\cdots\otimes\bold{1},$$ $$H_2=\bold{1} \otimes E_2\bold{1}\otimes\cdots\otimes\bold{1},$$ $$\dots$$ $$H_n=\bold{1}\otimes\bold{1}\otimes\cdots\otimes E_n\bold{1}$$
and the time evolution is a simple multiplying by an overall phase factor
\[
\ket{\psi(t)}=e^{-i(E_1+E_2+\cdots+E_n)(t-t_0)}\ket{\psi_0}\ ,
\]
where
\[
\ket{\psi_0}=\sum_{i_1,i_2...,i_n}\alpha_{i_1i_2...i_n}\ket{i_1}\ket{i_2}...\ket{i_n}\ .
\]

All operators, in particular projectors, commute with such hamiltonian and thus in such experiments it does not matter at all when the measurements on the exclusive parts of the system happened. Suppose we measure observable $A$ on the first particle and observable $B$ on the second. Observable $A$ acts as an identity at the rest of the system (particles $2,\dots,n$) and the same holds for the $B$. $A$ and $B$ thus commute. Probability of measuring value $a$ of an observable $A$ in a time $t_1$ given the measured value $b$ of an observable $B$ in a time $t_2$ remains the same if we choose any different times, for example $t_3$ and $t_4$ respectively. For simplicity suppose our initial time $t_0=0$.
\begin{equation*}
\begin{split}
P&(A=a(t_1)|B=b(t_2))\\
&=\bra{\psi_0}e^{iHt_2}P_{B=b}e^{iH(t_1-t_2)}P_{A=a}e^{-iH(t_1-t_2)}P_{B=b}e^{-iHt_2}\ket{\psi_0}\\
&=\bra{\psi_0}P_{B=b}P_{A=a}P_{B=b}\ket{\psi_0}\\
&=\bra{\psi_0}P_{A=a}P_{B=b}P_{B=b}\ket{\psi_0}\\
&=\bra{\psi_0}P_{A=a}P_{B=b}\ket{\psi_0}\\
&=\bra{\psi_0}P_{B=b}P_{A=a}\ket{\psi_0}\\
&=P(A=a(t_3)|B=b(t_4))
\end{split}
\end{equation*}

Now we see in such systems it does not matter when the system is measured, nor it does not matter on the time ordering of the measurements.

\chapter{List of unitary transformations used in the thesis}\label{listofUnitaries}
\section{Unitary 1}\label{unitary1}
States just before the interaction $\ket{\psi_u(0)}$, $\ket{\psi_d(0)}$ are orthogonal and if we suppose they are normalized too the unitary operator changing
\[
\frac{1}{\sqrt{2}}\ket{0}(\ket{\psi_u(0)}+\ket{\psi_d(0)})\ ,
\]
to
\[
\frac{1}{\sqrt{2}}(\ket{0}\ket{\psi_u(0)}+\ket{1}\ket{\psi_d(0)})\ ,
\]
is for example\footnote{Unitary transformation doing this is not unique.} of the form
\begin{equation*}
\left( \begin{array}{cccc}
1 & 0 & 0 & 0 \\
0 & 0 & 0 & 1 \\
0 & 0 & 1 & 0 \\
0 & 1 & 0 & 0 \end{array} \right),
\end{equation*}
in basis $\{\ket{0}\ket{\psi_u(0)},\ket{0}\ket{\psi_d(0)},\ket{1}\ket{\psi_u(0)},\ket{1}\ket{\psi_d(0)}\}$.

\section{Unitary 2}\label{unitary2}
The unitary matrix between vectors $\ket{i}=\left(\sqrt{\frac{2}{5}},\frac{1}{\sqrt{10}},\sqrt{\frac{2}{5}},\frac{1}{\sqrt{10}}\right)^T$ $\ket{f}=\left(\frac{1}{\sqrt{2}},0,\frac{3}{5\sqrt{2}},\frac{2\sqrt{2}}{5}\right)^T$ (in basis $\{\ket{0}\ket{L},\ket{0}\ket{R},\ket{1}\ket{L},\ket{1}\ket{R}\}$) is for example
\[
U=P_2P_1^{-1}=P_2P_1^T\ ,
\]
where
\begin{equation*}
P_1=\left( \begin{array}{cccc}
\sqrt{\frac{2}{5}} & \frac{1}{\sqrt{2}} & 0 & -\frac{1}{\sqrt{10}} \\
\frac{1}{\sqrt{10}} & 0 & \frac{1}{\sqrt{2}} & \sqrt{\frac{2}{5}} \\
\sqrt{\frac{2}{5}} & -\frac{1}{\sqrt{2}} & 0 & -\frac{1}{\sqrt{10}} \\
\frac{1}{\sqrt{10}} & 0 & -\frac{1}{\sqrt{2}} & \sqrt{\frac{2}{5}} \end{array} \right),\ \ \
P_2=\left( \begin{array}{cccc}
\frac{1}{\sqrt{2}} & 0 & \frac{3}{5\sqrt{2}} & \frac{2}{\sqrt{17}} \\
0 & 1 & 0 & 0 \\
\frac{3}{5\sqrt{2}} & 0 & -\frac{1}{\sqrt{2}} & \frac{6}{5\sqrt{17}} \\
\frac{2\sqrt{2}}{5} & 0 & 0 & -\frac{\sqrt{17}}{5} \end{array} \right)\ .
\end{equation*}

\section{Unitary 3}\label{unitary3}
\begin{equation*}
U_1=\left( \begin{array}{cccccc}
0 & 0 & 1 & 0 & 0 & 0 \\
0 & 0 & 0 & 0 & 0 & 1 \\
1 & 0 & 0 & 0 & 0 & 0 \\
0 & 0 & 0 & 0 & 1 & 0 \\
0 & 0 & 0 & 1 & 0 & 0 \\
0 & 1 & 0 & 0 & 0 & 0 \end{array} \right),
\end{equation*}
acting between Alice and Car in basis $\{0L,0R,1L,1R,2L,2R\}$ (schematically).

\begin{equation*}
U_1=\left( \begin{array}{cccccc}
0 & 0 & \frac{1}{\sqrt{2}} & \frac{1}{\sqrt{2}} & 0 & 0 \\
0 & 0 & 0 & 0 & \frac{1}{\sqrt{2}} & \frac{1}{\sqrt{2}} \\
0 & 0 & \frac{1}{\sqrt{2}} & -\frac{1}{\sqrt{2}} & 0 & 0 \\
0 & 0 & 0 & 0 & \frac{1}{\sqrt{2}} & -\frac{1}{\sqrt{2}} \\
1 & 0 & 0 & 0 & 0 & 0 \\
0 & 1 & 0 & 0 & 0 & 0 \end{array} \right),
\end{equation*}
acting between Alice and Switching unit in basis $\{0u,0d,1u,1d,2u,2d\}$.

\begin{equation*}
U_3=\frac{1}{\sqrt{2}}\left( \begin{array}{cccc}
1 & 1 & 0 & 0 \\
1 & -1 & 0 & 0 \\
0 & 0 & -1 & 1 \\
0 & 0 & 1 & 1 \end{array} \right),
\end{equation*}
acting between Car and Switching unit in basis $\{Lu,Ld,Ru,Rd\}$.

\section{Unitary 4}\label{unitary4}
\begin{equation*}
U=e^{-it}\left( \begin{array}{cccccc}
\cos t & 0 & i \sin t & 0 & 0 & 0 \\
0 & \cos t & 0 & i\sin t & 0 & 0 \\
i\sin t & 0 & \cos t & 0 & 0 & 0 \\
0 & i \sin t & 0 & \cos t & 0 & 0 \\
0 & 0 & 0 & 0 & 1 & 0 \\
0 & 0 & 0 & 0 & 0 & 1 \end{array} \right),
\end{equation*}
in basis $$\{\neutranie \text{alive},\neutranie \text{dead},\smiley \text{dead},\frownie \text{alive},\smiley \text{alive},\frownie \text{dead}\}$$ (schematically).

\chapter{Very short introductions to some interpretations of Quantum Mechanics}

\section{Everett's Many World interpretation}

Hugh Everett's Relative states formulation, later renamed by Bryce DeWitt as Many World interpretation is based on the idea whole quantum mechanics together with Born rule of probability can be derived using only unitary evolution. There is not general consensus whether it was successful \cite{LandsmanManyWorldBornRuleSuccesful}, \cite{KentManyWorldCritique}. Process of measurement in Many world interpretation is just entangling observer with the system from the point of view of observer.

Let the resulting state be for example
\[
\frac{1}{2}\ket{\text{dead cat}}\ket{\smiley}+\frac{\sqrt{3}}{2}\ket{\text{alive cat}}\ket{\frownie}\ .
\]
Many world interpretation says the world splits into two worlds. Observer enters world where cat dies with probability $(\frac{1}{2})^2=\frac{1}{4}$ and is smiling because he does not like cats and with probability $(\frac{\sqrt{3}}{2})^2=\frac{3}{4}$ observer enters world where the cat is still alive and he is happy.

\section{Decoherence theory}

Decoherence theory see the measurement as interacting of particle with the environment, which is complex system usually consisted of millions of particles. Still the environment can be described by some state (although we do not know it exactly) and interacting with the particle changes this state in some way. Put mathematically, let
\[
\ket{\tilde\psi_0}=\sum_i\ket{i}\braket{i}{\psi}\ ,
\]
where $\{\ket{i}\}$ is einselected basis (environmentally induced selected basis). Total wave function of particle plus the environment is
\[
\ket{\psi_0}=\sum_i\ket{i}\ket{\epsilon}\braket{i}{\psi}\ ,
\]
where $\ket{\epsilon}$ is the initial state of the environment. Each state $\ket{i}\ket{\epsilon}$ evolves to $\ket{\epsilon_i}$ through some unitary evolution. Thus the final state is
\[
\ket{\psi_0}=\sum_i\ket{\epsilon_i}\braket{i}{\psi}\ .
\]
Since unitary evolution conserves orthogonality, we have
\[
\braket{\epsilon_i}{\epsilon_j}=\braket{i}{j}=\delta_{i,j}
\]

We cannot effectively control all degrees of freedom of the environment and that is why we need to use the density matrix and trace over the environment to describe our particle. But now the particle is not anymore in a pure state, it is in a mixed state, which is known as decoherence. Eigenvectors of this new density matrix are exactly the states our former state $\ket{\tilde\psi_0}$ can pass onto which fits perfectly the well-known Born rule.

\section{Consistent Histories}

Consistent histories approach is based on the notion of Proposition, such as ``The particle went through upper slit at a time $t$''. Set of propositions form a history. \emph{Homogenous history} $H_i$ is a sequence of propositions $P_{i,j}$, where index $j$ refers to time.
\[
H_i=(P_{i,1},P_{i,1},\dots P_{i,n_i})
\]
meaning proposition $P_{i,1}$ is true at time $t_1$, then proposition $P_{i,2}$ is true at time $t_2$ etc.

Each proposition can be represented by a projection operator $\hat P_{i,j}$ acting on a Hilbert space. Homogenous history is represented by tensor product of propositions
\[
\hat H_i=\hat P_{i,1}\otimes\hat P_{i,2}\otimes\dots\otimes\hat P_{i,n_i}
\]

We define \emph{Class operator} acting on a history as
\[
\hat C_{\hat H_i}=T\prod_{j=1}^{n_i}P_{i,j}=\hat P_{i,1}\hat P_{i,2}\dots\hat P_{i,n_i} ,
\]
which orders Projectors $\hat P_{i,j}$ chronologically, i.e. $t_j\geq t_{j+1}$ .

Set of Histories $\{H_i\}$ is \emph{consistent} if
\[
\Tr{\hat C_{\hat H_i}\rho\hat C_{\hat H_j}^{\dagger}}=0
\]
for all $i\neq j$ and initial density operator $\rho$.

Probabilities of history $\hat H_i$ is then
\[
Pr(\hat H_i)=\Tr{\hat C_{\hat H_i}\rho\hat C_{\hat H_i}^{\dagger}}\ .
\]

\bibliographystyle{plain}
\bibliography{myrefs1}

\begin{thebibliography}{10}

\bibitem{AharonovCanaFutureChoiceAffect}
Y.~Aharonov, E.~Cohen, D.~Grossman, and A.~C. Elitzur.
\newblock {Can a Future Choice Affect a Past Measurement's Outcome?}, August
  2012.
\newblock [quant-ph/1206.6224].

\bibitem{MultiparticleInterferometryGreenbergerZeilinger}
A.~Zeilinger D.~M.~Greenberger, M. A.~Horne.
\newblock {Mathematica Foundations of Quantum Theory}.
\newblock {\em Physics Today}, 46:22--29, 1993.

\bibitem{DemystifyingTheDelayedChoice}
B.~Gaasbeek.
\newblock {Demystifying the Delayed Choice Experiments}.
\newblock 2010.
\newblock [quant-ph/1007.3977v1].

\bibitem{ComplementarityHerzog}
T.~J. Herzog, P.~G. Kwiat, H.~Weinfurter, and A.~Zeilinger.
\newblock {Complementarity and the Quantum Eraser}.
\newblock {\em Phys. Rev. Lett.}, 75:3034, 1995.

\bibitem{ItanoZenoEffectExperimental}
Wayne~M. Itano, D.~J. Heinzen, J.~J. Bollinger, and D.~J. Wineland.
\newblock {Quantum Zeno effect}.
\newblock {\em Physical Review A}, 41:2295--2300, March 1990.

\bibitem{JaquesExperimentalRealizationDelayedChoice}
V.~Jacques, E.~Wu, F.~Grosshans, F.~Treussart, P.~Grangier, A.~Aspect, and
  J.~Roch.
\newblock {Experimental Realization of Wheeler's Delayed-Choice Gedanken
  Experiment}.
\newblock In {\em {Conference on Coherence and Quantum Optics}}, page CWB4.
  Optical Society of America, 2007.

\bibitem{KaiserDelayedChoiceExperimental}
F.~Kaiser, T.~Coudreau, P.~Milman, D.~B. Ostrowsky, and S.~Tanzilli.
\newblock {Entanglement-Enabled Delayed-Choice Experiment}.
\newblock {\em Science}, 338(6107):637--640, November 2012.

\bibitem{KentManyWorldCritique}
A.~Kent.
\newblock {One world versus many: the inadequacy of Everettian accounts of
  evolution, probability, and scientific confirmation}, August 2010.

\bibitem{KimDelayedChoiceQuantumEraser}
Y.~Kim, R.~Yu, S.~P. Kulik, Y.~Shih, and M.~O. Scully.
\newblock {Delayed Choice Quantum Eraser}.
\newblock {\em Phys. Rev. Lett.}, 84:1--5, 2000.

\bibitem{PatiNoDeleting}
A.~Kumar~Pati and S.~L. Braunstein.
\newblock {Impossibility of deleting an unknown quantum state}.
\newblock {\em Nature}, 404(6774):164--165, March 2000.

\bibitem{LandsmanManyWorldBornRuleSuccesful}
N.P. Landsman.
\newblock {The conclusion seems to be that no generally accepted derivation of
  the Born rule has been given to date, but this does not imply that such a
  derivation is impossible in principle.}
\newblock In {\em {Compendium of Quantum Physics (eds.) F.Weinert, K.
  Hentschel, D.Greenberger and B. Falkenburg}}. Springer, 2008.

\bibitem{GagenContinuousPositionMeasurements}
G.~J.~Milburn M.~J.~Gagen, H. M.~Wiseman.
\newblock {Continuous position measurements and the quantum Zeno effect}.
\newblock {\em Physical Review A}, 48:132--142, 1993.

\bibitem{ZeilingerDelayedChoiceExperimentalSwapping}
Xiao-song Ma, S.~Zotter, J.~Kofler, R.~Ursin, T.~Jennewein, \v{C}. Brukner, and
  A.~Zeilinger.
\newblock {Experimental delayed-choice entanglement swapping}, October 2012.
\newblock [quant-ph/1203.4834].

\bibitem{NielsenChuang}
M.~A. Nielsen and I.~L. Chuang.
\newblock {\em {Quantum Computation and Quantum Information (Cambridge Series
  on Information and the Natural Sciences)}}.
\newblock Cambridge University Press, 2004.

\bibitem{PalssonViolatingBellExperimental}
M.~S. Palsson, J.~J. Wallman, A.~J. Bennet, and G.~J. Pryde.
\newblock {Experimentally Violating Bell Inequalities Without Complete
  Reference Frames}.
\newblock March 2012.
\newblock [quant-ph/1203.6692].

\bibitem{PeresDelayedChoiceEntanglementSwapping}
A.~Peres.
\newblock {Delayed choice for entanglement swapping}.
\newblock {\em Journal of Modern Optics}, 47(2):139--143, 2000.

\bibitem{PeruzzoDelayedChoiceExperimental}
A.~Peruzzo, P.~J. Shadbolt, N.~Brunner, S.~Popescu, and J.~L. O'Brien.
\newblock {A quantum delayed choice experiment}, June 2012.
\newblock [quant-ph/1205.4926].

\bibitem{SchlosshauerDecoherenceMeasurementInterpretations}
M.~Schlosshauer.
\newblock {Decoherence, the measurement problem, and interpretations of quantum
  mechanics}.
\newblock {\em Reviews of Modern Physics}, 76(4):1267--1305, February 2005.

\bibitem{WheelersDelayedChoice}
J.~A. Wheeler.
\newblock {\em {Mathematica Foundations of Quantum Theory}}, chapter {The Past
  and Delayed Choice Double Slit Experiment}.
\newblock Academic Press, 1978.

\end{thebibliography}
\addcontentsline{toc}{chapter}{Bibliography}
\end{document}